\documentclass{sig-alternate}
\usepackage{mathptmx} 
\usepackage{fancyhdr}
\usepackage[normalem]{ulem}
\usepackage[hyphens]{url}
\usepackage[sort,nocompress]{cite}
\usepackage[final]{microtype}
\usepackage[keeplastbox]{flushend}
\usepackage{microtype}
\usepackage{graphicx}
\usepackage{subcaption}
\usepackage[font={normalsize,bf}]{caption}
\usepackage{booktabs} 
\usepackage{url}
\usepackage{xspace}
\usepackage{amsmath,amssymb,amsfonts}
\usepackage{paralist}
\usepackage{bm} 
\usepackage{xcolor} 
\usepackage{algorithm}
\usepackage{algorithmic}
\usepackage[bookmarks=true,pdfpagelabels=false,breaklinks=true,colorlinks,linkcolor=black,citecolor=blue,urlcolor=black]{hyperref}
\usepackage{setspace}
\usepackage[capitalise,nameinlink]{cleveref}
\Crefformat{figure}{#2Fig.~#1#3}
\Crefformat{table}{#2Tab.~#1#3}
\Crefformat{section}{#2Sec.~#1#3}

\newcommand{\etal}{\textit{et al.}\xspace}

\newcommand{\cbpfive}{CBP-5\xspace}

\newcommand{\specint}{SPECINT2017\xspace}

\newcommand{\tage}{{\sc TAGE}\xspace}
\newcommand{\tagescl}{{\sc \tage-SC-L}\xspace}
\newcommand{\tagesclBig}{{\sc \tagescl 64KB}\xspace}

\newcommand{\tagesclSmall}{{\sc \tagescl 8KB}\xspace}

\newcommand{\SLBIU}{{\sc SLBIU}\xspace}
\newcommand{\numlocalhistbits}{lh}
\newcommand{\numglobalhistbits}{gh}     
\newcommand{\numhistbits}{l}   
\newcommand{\numoffloadedbranches}{n}       
\newcommand{\nnz}{nnz}
\newcommand{\numweightbits}{q}
\newcommand{\numpcbits}{p}
\newcommand{\HistorySelect}{{\sc History-Select($\nnz,\numhistbits$)}\xspace}
\newcommand{\SignFlip}{{\sc Sign-Flip($\nnz,\numweightbits$)}\xspace}

\newcommand{\loss}[1]{\mathrm{L}\left(#1\right) }
\DeclareMathAlphabet{\pazocal}{OMS}{zplm}{m}{n}

\newcommand{\onenorm}[1]{\ensuremath{\left\|#1\right\|_1}}

\newcommand{\RR}{\mathbb{R}}

\newcommand{\x}{\bm{x}}
\newcommand{\w}{\bm{w}}

\newcommand{\lassohints}{sparsity hints\xspace}
\newcommand{\Lassohints}{Sparsity hints\xspace}

\newcommand{\hint}{f_{\tiny \text{\branch}}}

\newcommand{\branch}{pc}

\newcommand{\score}{S}

\newlength\myindent
\title{Identifying and Exploiting Sparse Branch Correlations for
Optimizing Branch Prediction} 
\author{
Anastasios Zouzias\textsuperscript{*}, Kleovoulos Kalaitzidis\textsuperscript{*}, Konstantin Berestizshevsky\textsuperscript{\textdagger}, \\Renzo Andri\textsuperscript{*}, Leeor Peled\textsuperscript{\textdaggerdbl}, Zhe Wang\textsuperscript{*}\\ 
*~Zurich Research Center, Huawei Technologies Switzerland \\
\textdagger~School of Electrical Engineering, Tel Aviv University Israel\\
\textdaggerdbl~Tel-Aviv Research Center, Huawei Technologies Israel
}

\begin{document}
\maketitle
\pagestyle{plain}

\begin{abstract}
Branch prediction is arguably one of the most important speculative mechanisms within a high-performance processor architecture. A common approach to improve branch prediction accuracy is to employ lengthy history records of previously seen branch directions to capture distant correlations between branches. The larger the history, the richer the information that the predictor can exploit for discovering predictive patterns. However, without appropriate filtering, such an approach may also heavily disorganize the predictor's internal mechanisms, leading to diminishing returns. This paper studies a fundamental control-flow property: the sparsity in the correlation between branches and recent history. First, we show that sparse branch correlations exist in standard applications and, more importantly, such correlations can be computed efficiently using sparse modeling methods. Second, we introduce a sparsity-aware branch prediction mechanism that can compactly encode and store sparse models to unlock essential performance opportunities. We evaluated our approach for various design parameters demonstrating MPKI improvements of up to 42\% (2.3\% on average) with 2KB of additional storage overhead. Our circuit-level evaluation of the design showed that it can operate within accepted branch prediction latencies, and under reasonable power and area limitations.
\end{abstract}
%
\section{Introduction \& Motivation}\label{sec:intro}
%
Superscalar pipelined processors rely profoundly on speculative and out-of-order (OoO) execution to keep their pipeline busy, hiding latency with instruction- and memory-level parallelism (ILP and MLP). Branch prediction (BP) is the key mechanism that drives speculative execution by steering the front end of the pipeline, each time predicting the instructions that follow in the stream. Working on the correct code path is paramount to performance since recovering from a branch misprediction and refilling the processor's instruction reservoirs (ROB, instruction/issue queue, etc.) after a pipeline restart can cost dozens of cycles~\cite{z15_IBM}. Notably, as the misprediction penalty gets larger due to the increasing pipeline depth and width~\cite{Sprangle2002, michaud2001, karkhanis2004}, modern processors have come to rely on incredibly accurate branch predictors.
%

%
\begin{figure}[!t]
\centering
\includegraphics[width=0.475\textwidth]{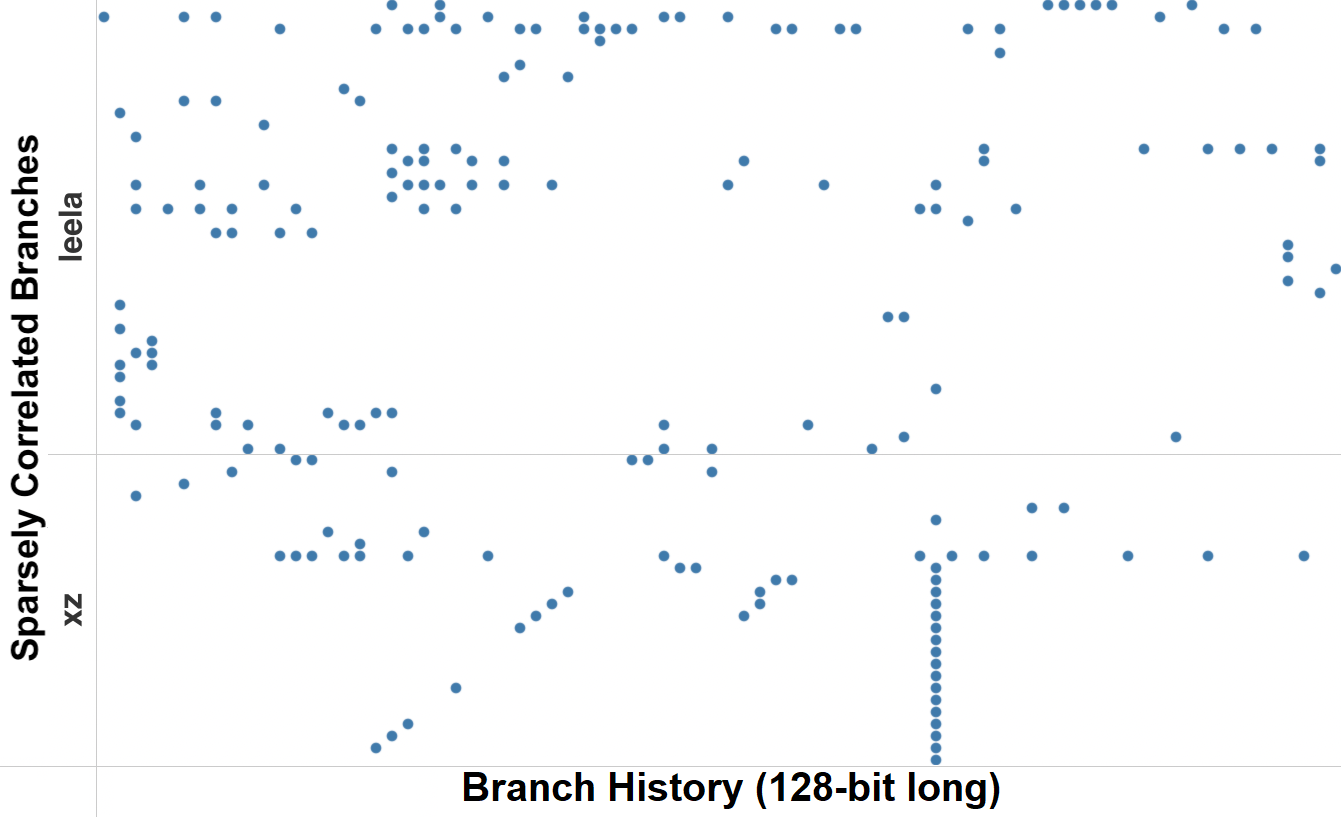}
\caption{Sparse branch history correlation in two of the \specint benchmarks with the highest MPKI.}
\label{fig:main}
\end{figure}%
%
%

%
Today, virtually all branch predictors in high-end processors are highly-engineered variants of TAGE~\cite{seznec2006} and Perceptron~\cite{perceptron2001}. Over the past 15 years, branch prediction has been primarily focused on these two designs, resulting in some highly sophisticated branch predictors that can overall achieve remarkably high prediction accuracy~\cite{seznec2016, jimenez2016}. However, recent studies demonstrate that the missing accuracy gap limits substantially the performance scalability of future processors\cite{bp:not_solved:2019, bp:auto-predication-2020}. As the further enhancement of BP methods is becoming increasingly challenging, performance improvements have stagnated the recent years. Lately, the focus has been concentrated on improving the prediction accuracy of a specific class of branches that are regularly mispredicted. In this work, we argue that there are still uncapped opportunities that can unveil key improvements in branch prediction not by strictly focusing on what is classified as {\it ``hard-to-predict''} (H2P), but by exploring vital control-flow properties that have been systematically overlooked.
More specifically, despite extensive differences in their prediction mechanisms, state-of-the-art branch predictors learn correlations between branches through the use of large hardware tables and long histories of previous branch outcomes (taken/not-taken), with efficacy strongly associated with the amount of storage available. An essential observation made from both TAGE- and Perceptron-based predictors is that predictive signatures in branch history occur with varying lengths. That is, long histories may be able to expose important branch patterns that shorter ones fail to and vice versa. As such, modern BP designs are built with a set of input features that are based on branch histories with different lengths (in number of bits)~\cite{isca:exynos,z15_IBM}.
Ultimately, longer histories can elicit correlations with the more distant past, but in the general case, the correlation of a branch with a large history is rather sparse, i.e., only a few selective parts of the history are eventually informative of the branch outcome~\cite{evers:phdthesis,evers:branch_sparsity}. As an example, in \cref{fig:main}, we demonstrate the sparsity of branch correlations in two \specint applications (\texttt{xz} and \texttt{leela}) with the highest branch misprediction rate~\cite{limaye2018workload}. To characterize sparsity, we use linear regression analysis methods. In particular, we perform Lasso logistic regression~\cite{book:stat_learning_lasso} over all the static\footnote{A static instruction (either branch or not) is uniquely identified by its PC (program counter) address and might be executed dynamically more than once for a specific program execution, e.g., in a loop.} branches in our dataset, and we build sparse linear functions (models) that predict branches based on the branch-outcome history ($128$-bit long). We choose to employ linear models as they lend themselves well to a pragmatic hardware implementation. We only consider sparse models of over 99\% accuracy that relate to branches that are not highly biased towards a specific direction. Rows in \cref{fig:main} correspond to a few such branches per application, while each mark indicates the existence of correlation between their outcome and the corresponding history location. 
As illustrated, the screened branches are correlated with at most 20 sparse history locations, and in some cases, only with a single one. Nonetheless, modern history-based branch predictors do not employ methods for discarding the non-informative (noisy) parts of the branch history during prediction (TAGE-based) or they attempt to identify them online (i.e., at run-time) with weight balancing (Perceptron-based). This fundamental design decision leads to a catastrophic explosion of the number of table entries required for tracking all the observable history patterns since such number grows exponentially with the history length~\cite{fern2000dynamic}. As branch predictors are traditionally anticipated to be trained online, designers resort to various heuristics trying to workaround this issue. Yet, the focus has been mainly on achieving a high enough accuracy per bit of storage capacity rather than on revising rigorously the key prediction principles~\cite{amd-zen-2}.
Our work aims to fill this gap by attending to only the few important bits in the branch history vectors, while discarding the non-informative ones. To properly filter the branch history at this granularity, it is necessary to adopt more detailed training algorithms through additional compiler support, otherwise dubbed {\it``offline training''}. Fortunately, in the era where data center/cloud applications surge, the otherwise overwhelming cost of offline training can be amortized through economies of scale. In this work, we employ an \textit{``offline-training/online-inference''} paradigm, as detailed by Lin \& Tarsa~\cite{bp:not_solved:2019}, for introducing \textit{sparsity-aware branch prediction}.
Specifically, our contributions are as follows:
\begin{enumerate}
    \item \textbf{Effective detection of sparsely correlated branches}. We demonstrate that in practical workloads there exist branches whose direction (i.e. taken/not-taken) can be modeled with a sparse linear mapping from their history. In particular, we employ an offline methodology for determining sparse linear model parameters per branch that we call ``sparsity hints''.
    \item \textbf{Novel sparsity-aware branch prediction design}. We show that sparsity hints \textit{improve prediction accuracy} using a \textit{compact storage}. We present a detailed microarchitectural design of a branch predictor, dubbed {\it Sparse Linear Branch Inference Unit} (\SLBIU), that uses sparsity hints for a few selected branches, operating as an auxiliary component alongside a primary branch predictor. Overall, our scheme improves the prediction accuracy on \cbpfive~\cite{cbp2016} and \specint~\cite{spec2017} benchmark suites, compared to \tagesclSmall with $2$KB of storage overhead. 
    Moreover, our design can operate within the latency acceptable by the CPU front-end for branch prediction, while remaining under reasonable area and power limitations.
\end{enumerate}
%

%
\begin{table*}[ht]
\scriptsize
\begin{center}
  \begin{tabular}{lcr p{0.01\textwidth} rr p{0.01\textwidth} cc}
	\toprule
  &  \multicolumn{2}{c}{\footnotesize Oracle Sparse Prediction}  && \multicolumn{2}{c}{\footnotesize TAGE-SC-L Misses} && \multicolumn{2}{c}{\footnotesize TAGE-SCL-L Entries (avg)  | Allocations} 
  \\ 
  \cmidrule{2-3}\cmidrule{5-6}\cmidrule{8-9}

 {\footnotesize Trace / branch} &  \texttt{\footnotesize History bits} & \texttt{\footnotesize  Misses} && \texttt{\footnotesize 8KB}& \texttt{\footnotesize  64KB} && \texttt{\footnotesize  8KB} & \texttt{\footnotesize 64KB}   \\  
\cmidrule{2-3}\cmidrule{5-6}\cmidrule{8-9}
 {\scriptsize LONG-MOBILE-1 / 548221168352}  & 1   & 6,118  && \textbf{657,293}  & 12,734 && 443.5 | \textbf{549K} & \textbf{1,131.9} | 47K \\
  {\scriptsize SHORT-MOBILE-16 / 1566871128} & 7   & 3,697  && \textbf{194,634}  & 10,181 && 36.8 | \textbf{328K} & \textbf{150.9} | 24K \\
  {\scriptsize SHORT-SERVER-225 / 5564716}   & 1   & 45,794 && 103,383 & 74,393 && \textbf{311.7} | 56K & \textbf{3,022.14} | 83K \\
  {\scriptsize SHORT-MOBILE-60 / 50044 }     & 7   &	711 &&	66,516 &	54,686 &&	61.9 | 72K &	\textbf{601.9} | 75K \\
  {\scriptsize LONG-MOBILE-24 / 50044}       & 7   &	711 &&	66,310 &	54,976 &&	69.3 | 71K &	\textbf{834.5} | 77K \\
  {\scriptsize SHORT-MOBILE-59 / 50044}      & 9   &		879	 &&	64,126 &		33,102 &&	73.5 | 40K &	\textbf{421.7} | 42K \\
    \bottomrule
    \end{tabular}
\end{center}
  \caption{Motivating branches from the \cbpfive trace set that are sparsely correlated with a 1024-bit history (global and local concatenated) based on sparse models of $\geq 0.99$ accuracy. Rows correspond to statistics (mispredictions, unique entries and allocations) per static branch, sorted by TAGE-SC-L 8KB mispredictions. Unique entries are taken as the average from periodical snapshots over execution.}
  \label{tab:motivating_examples}
\end{table*}
%
%
\section{Current predictors Limitations}\label{sec:sota_limits}
%
State-of-the-art branch predictors, such as \tage~\cite{seznec2006} and Perceptron~\cite{perceptron2001}, are tabular and history-based, i.e., they identify predictable branch patterns by mapping recent branch-outcome histories to internal state machines (saturating counters or weights) stored in table-entries. Most commonly, they use various formations of the local (previous outcomes of a static branch) and the global history record (prior outcomes of any branch) and the branch program counter (PC). Intuitively, branch histories are expressed as a sequence of consecutive binary events (\emph{taken/not-taken}) leading to a branch. For example, the global history is a $k$-bit sequence representing the $k$ branch directions before a certain branch. As such, history-based predictors attempt to identify predictable patterns in the form of thick series of events, even when branches correlate sparsely with them. Ideally, a separate table-entry will be allocated for each observable pattern. As the number of patterns grows exponentially with the history length, the required number of entries becomes quickly too large. 
Current branch predictors use astonishingly large branch histories. TAGE-SC-L~\cite{seznec2016}, the most recent TAGE variant and winner of the last BP championship (\cbpfive~\cite{cbp2016}), tracks histories with lengths up to $3,000$ bits, whereas the Multiperspective Perceptron predictor~\cite{jimenez2016} (ranked second in \cbpfive), uses several features acquired by tracking similarly sized histories. As the employed histories become longer, more previous branch directions can be looked up, and thus, the chances of identifying correlation with more distant branches increases. Nonetheless, so does the amount of sparsity and variation of the predictive signatures.
Consider the simple case of two $if$-type branches $A$ and $B$ with dependable conditions, where branch $A$ precedes branch $B$ in the program order. In such a scenario, the outcome of $B$ correlates with the one of $A$. If there are $M$ other $if$-type branches between $A$ and $B$ with independent conditions that change directions constantly, there will be at least $2^{(M+1)}$ different patterns that the predictor needs to distinguish to predict $B$ accurately. These intermediate $M$ branch-outcomes can be considered as {\it "noise"} in the history, since $B$ can be predicted accurately by solely monitoring $A$. This noise burdens the predictor with superfluous increase of entries allocations for tracking all the observable patterns. Under storage constraints, such allocation pressure can heavily disorganize the predictor's state machines, compromising accuracy. Although branch predictors employ various storage-saving heuristics to generally mitigate entries allocation, they miss sparsity by tracking history patterns in a quite dense form.
TAGE predictors are based on the PPM data compression scheme~\cite{ppm1984}. They employ a plurality of tables that are indexed using overlapping history slices of increasing lengths, hashed (XOR-ed) with the branch address. Table entries contain (partially) tagged saturating counters that model the prediction. The matching entry that is accessed with the longest history provides eventually the prediction. As in PPM, the intuition is that longer histories should be used when shorter ones fail to be accurate. Yet, even when the correlation of a branch with the history is sparse, TAGE predictors need to allocate storage for tracking all the possible patterns, since the histories are used in a compact way, i.e., without an explicit mechanism for disregarding the non-informative parts. Glimpsing at \cref{tab:motivating_examples} (detailed in \cref{sec:bp_lasso}) demonstrates the effects of the excessive entries allocation when histories are noisy. Especially under the storage pressure of 8KB, TAGE-SC-L repeatedly "forgets" and "relearns" branch patterns, inducing a much higher number of mispredictions than at 64KB.
Perceptron predictors are loosely based on the homonymous learning algorithm. Their prediction mechanism receives a vector of inputs with corresponding weights and computes their dot product, which is then thresholded to provide the prediction. Input weights are trained and adapted according to the correctness of the produced predictions. Initial proposals used only the global history~\cite{perceptron2001}, while in the latest Perceptron predictors, the input vector consists of several different hashes (organizations) of the branch history~\cite{tarjan2005, jimenez2016}. Still, without filtering the branch history a priori, Perceptron suffers from significant aliasing among noisy histories and their synthetic formations.
As in our dataset (see \cref{sec:evaluation_methodology:setup}), Perceptron predictors achieve lower accuracy than TAGE predictors, we employ the latest TAGE-SC-L models~\cite{seznec2016} for our study (explained in \cref{sec:evaluation_methodology:setup}). In next section we explore the application of sparse modeling in BP for capturing the sparse correlations between branches and recent history. We specifically focus on sparse linear models, dropping the non-linear ones that can induce excessive storage and computational overheads. As we will show, sparse linear modeling effectively captures branches' correlation with a commonly used set of features, namely the global and local branch history, and inherently identifies its sparsity. 
%

%
\section{Sparse Branch Correlations}\label{sec:bp_lasso}
%
We start our exploration by outlining in \cref{tab:motivating_examples} a few examples that demonstrate the opportunity behind sparse branch-history correlations. Interestingly, the reported branch examples are not ``hard to predict'' for TAGE-SC-L (around a 1\% misprediction ratio). However, the average number of entries that TAGE-SC-L allocates for these branches is far from optimal since it has to replicate them exponentially per every noisy uncorrelated event (branch outcome) in the history. For example, \tagesclBig allocates 1K entries on average for a branch that could be predicted accurately with a single sparse signature (LONG-MOBILE-1 / 548221168352). \tagesclSmall requires $443.5$ entries on average for the same branch and, more importantly, induces around 51$\times{}$ mispredictions by not being able to suppress the effects of destructive aliasing in such a limited storage budget.
This storage efficiency issue is in line with previous work that exposed it by examining \tagesclBig with large code-footprint applications~\cite{bp:not_solved:2019}. Storing only sparse correlations could therefore generally improve the performance of state-of-the-art branch predictors by eliminating the need to represent irrelevant features in their storage, thereby reducing the predictor's footprint. Nonetheless, identifying and exploiting sparsity effectively is a grand challenge for branch predictors. By historically performing training solely online, typical branch-prediction designs are quite cumbersome for applying powerful sparse modeling. The next section provides a brief background of sparse linear modeling, revealing its interconnection with branch prediction. 
%
\subsection{Sparse Linear Modeling}\label{sec:background}
%
In supervised learning prediction tasks, often, only a small subset of input variables contributes to the prediction outcome. In the case of linear models where we focus, the problem is well-studied in literature with various synonyms, i.e., sparse modeling, sparse signal recovery, and Lasso regression~\cite{book:stat_learning_lasso}. In our study, we apply these techniques in the context of BP. Due to the binary nature of branch outcomes, BP resembles a classification problem. Thus, we opt to employ the Lasso logistic regression model, i.e., $\ell_1$ regularized logistic regression. Our main focus is on offline linear methods to understand the limitations and opportunities that sparsity could present in BP. However, in~\cref{sec:online_lasso}, we also examine the promising yet challenging case of online sparse linear modeling.
Assume a collection of input feature-vectors and target pairs: $(\x_1, y_1), (\x_2, y_2),\dots, (\x_m,y_m)$ where $\x_i\in{\RR}^l$ is the $i$-th input feature-vector and $y_i\in{\{0, 1\}}$ is the corresponding target. Lasso logistic regression seeks for a linear logistic function $f(\x):=\sigma(b+ \w^\top \x)$ where $\sigma(a):=1/(1+\exp(-a))$ is the sigmoid function, $b$ is the model bias/intercept, $\w^\top \x$ denotes the dot-product and $\w$ is the vector of model weights. The parameters $(\w, b)$ that define $f(\x)$, are determined by optimizing the following objective:
%
\begin{equation}\label{eqn:glm_l1_reg}
    \min_{b\in\RR, \w\in{\RR^l}} \left\{ \frac1{m} \sum_{i=1}^{m} \loss{y_i, f(\x_i)} + \lambda \onenorm{\w} \right\}
\end{equation}
%
for some $\lambda \geq 0$, $\loss{a,b}$ being the logistic loss and $\onenorm{\w}=\sum_{j=1}^{l} |w_j|$ the $\ell_1$-norm. The hyperparameter $\lambda$ enforces a trade-off between the model's sparsity and prediction performance. That is, in the special case where $\lambda=0$, Equation~\eqref{eqn:glm_l1_reg} corresponds to logistic regression (no model sparsity enforced). On the other extreme, it can be shown that $\w$ is the all-zeroes vector for a large enough value of $\lambda$. Setting an appropriate $\lambda$ value is not a trivial task, although methods such as coordinate descent~\cite{lasso:glmnet} can compute the full regularization path of $\lambda$ values. Such methods are also guaranteed to converge to a globally optimal solution~\cite{lasso:cd:tseng_2001}. In our experiments, we observed that a binary search approach on the range of $\lambda\in{[10^{-4},1.0]}$ with a stopping criterion of 99\% accuracy allows us to fine-tune $\lambda$ efficiently. Finally, we should note that there are several methods that can compute an optimal solution of Equation~\eqref{eqn:glm_l1_reg}~\cite{book:stat_learning_lasso}.
%

%
\subsection{Sparse Linear Models on Branch History}\label{sec:bp_lasso:sparse_modeling}
%

%
Branch predictors correlate the behavior of branches with the recent history of previous branch outcomes, expressed, most commonly, through the global and the local history records (GHR/LHR). To leverage the sparsity of branches' correlations, we construct sparse linear models that can express branch outcomes based on the branch history.
In particular, we aim to define the parameters (bias and sparse weights) of a logistic regression model $\hint(\x):=\sigma(b+w_1x_1+\dots + w_l x_l)$ so that $\hint(\x)\approx y$, where $\x$ is the branch history with branch directions represented as $\{\pm 1\}$, and $y$ is the modeled branch outcome. 
$\hint(\x)$ returns the prediction probability that the branch identified with  its \textit{PC} will be taken. To simplify the inference, we replace the sigmoid function with a sign check. If the sign is negative, the branch outcome is predicted not-taken and taken otherwise.
During training, once the history is sufficiently populated, we collect training samples in the form of $(\x_i,y_i)$ where $\x_i$ is the history vector and $y_i$ is the sampled branch outcome. Equipped with the above modeling configuration, Equation~\eqref{eqn:glm_l1_reg} can be optimized for every static branch. After training, based (mainly) on the accuracy and the sparsity of the resulting models of all the screened branches, it is defined which of them follows a sparse linear model. \cref{sec:approach:offline_process:selection} makes an in-depth analysis of this process.
%

%
\begin{figure}[!t]
\centering
\includegraphics[width=0.46\textwidth]{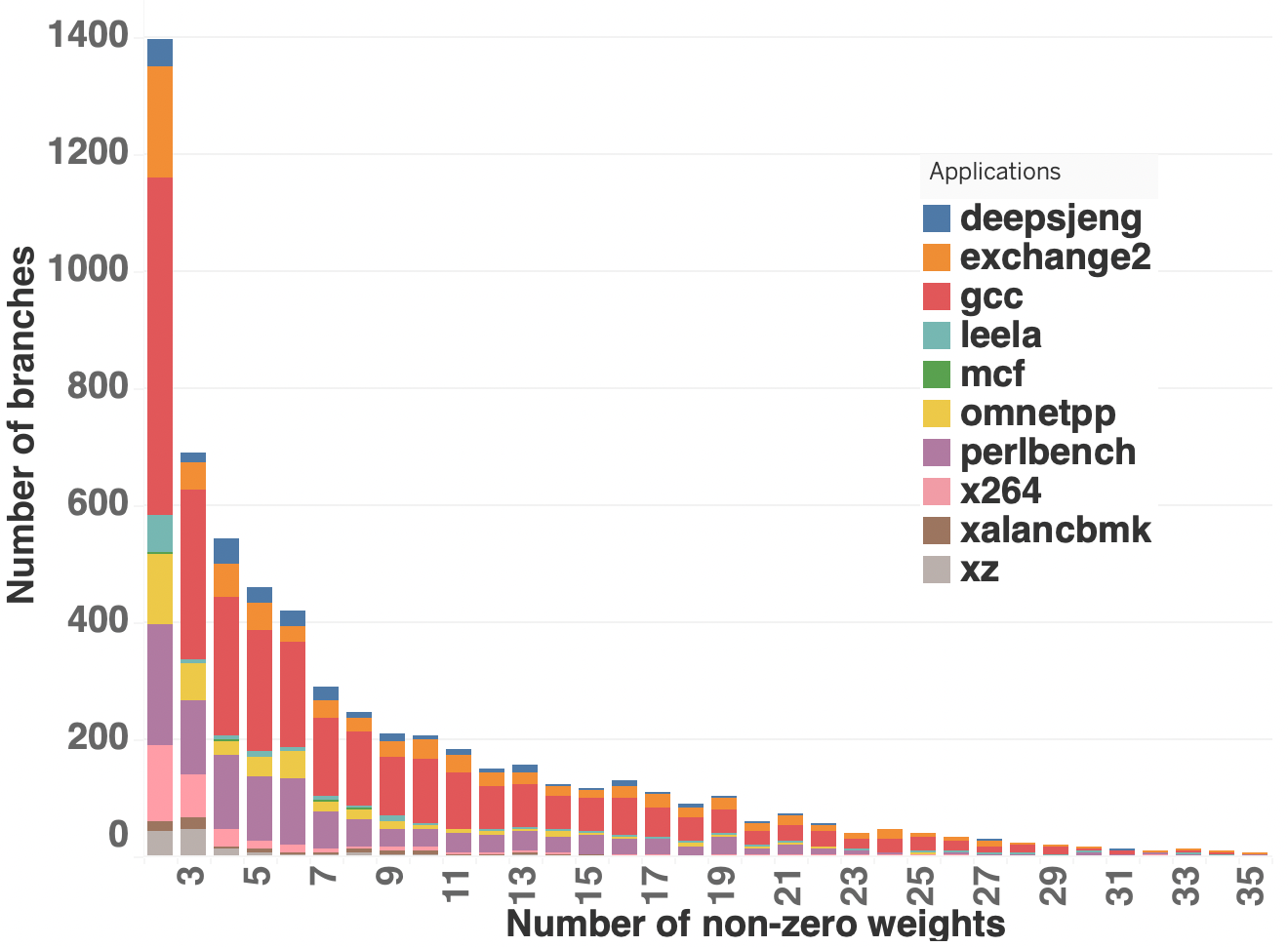}
\caption{Distribution based on non-zero weights.}
\label{fig:lasso_nnz_dist}
\end{figure}%
%
%
\begin{figure}[!t]
\center
\includegraphics[width=0.475\textwidth]{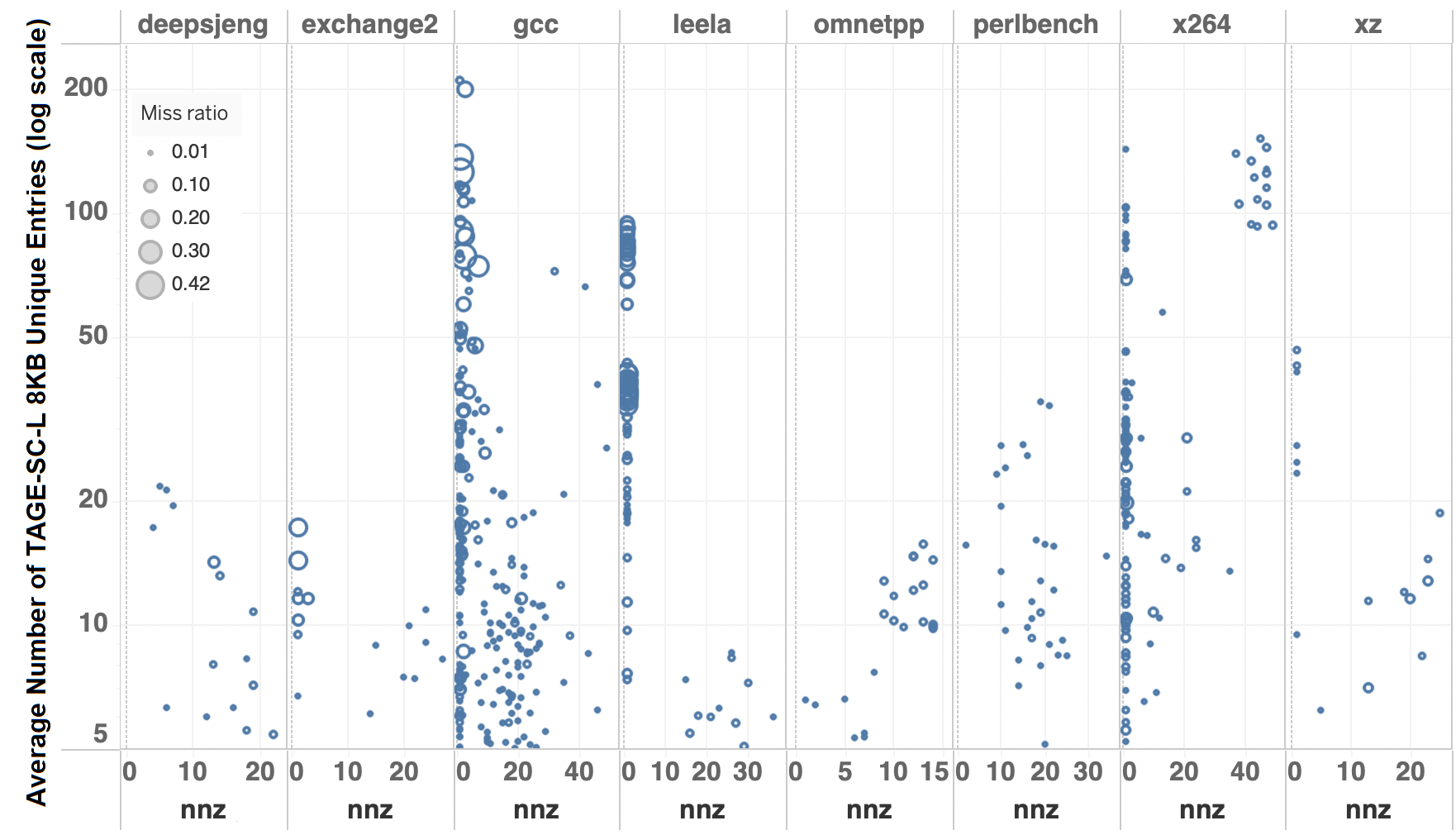}
\caption{Average entries vs non-zero weights.}
\label{fig:lasso_bias_vs_acc}
\end{figure}%
%
Based on this setup, we now perform sparsity analysis on traces of \specint \cite{spec2017} applications described in \cref{sec:evaluation_methodology:setup}. The branch history we use is the concatenation of a $512$-bit GHR and a $512$-bit LHR, i.e., $l:=1024$ and $\x := [\text{ghr}_1, \dots, \text{ghr}_{512},$ $\text{lhr}_1,\dots, \text{lhr}_{512}]$. We perform Lasso logistic regression on all the branches of each trace (with a few exceptions) to show that sparse linear correlations can be efficiently modeled. To strengthen our confidence in the statistical properties of the computed models, we exclude branches that are \emph{highly biased} towards a single direction, i.e., those with a taken rate less than $2\%$ or greater than $98\%$, and branches that appear less than $10K$ times in a trace. As the crux of our work is to design effective sparsity-aware branch predictors, we also set a rigid 99\% accuracy threshold, only above which, a sparse model is considered sufficiently accurate. In \cref{fig:lasso_nnz_dist} we plot the distribution of all the sufficiently accurate and non highly-biased  branches according to the number of the non-zero (nnz) weights of their model, as returned by sparse modeling. Note that per application, some branches may be counted more than once if they appear in multiple traces of the application.
As illustrated, most branches have a sharp decaying distribution based on the number of their non-zero (nnz) weights. Furthermore, the majority of them requires a fairly small number of non-zero weights in proportion to the employed history length ($1024$-bit long), bounded at $35$. Consider that for clarity we have also dropped the first histogram bin grouping branches of one single non-zero weight, as it was fairly large ($2,400$) and less challenging, representing essentially branches that correspond to loops.
According to our analysis, only a relatively small number of branches are identified as the interestingly sparse and accurate cases. 
Nonetheless, they can significantly affect the predictor's effectiveness by creating an increased allocation pressure when the available storage is limited. In our trace set, \tagesclSmall is plagued largely by such cases. In~\cref{fig:lasso_bias_vs_acc} we show numerous cases of branches\footnote{For clarity, in~\cref{fig:lasso_bias_vs_acc} we do not include \texttt{xalancbmk} and \texttt{mcf}, where we found very few branches with highly-accurate sparse models, as shown by \cref{fig:lasso_nnz_dist}.} whose sparse models contain only up to $20$ non-zero weights, i.e., those branches are correlated only with $20$ previous directions from the history, and still, they account for a massive amount of average unique entries in \tagesclSmall. Even more interestingly, \tagesclSmall mispredicts with a relatively high ratio a large fraction of these branches, yet, by dedicating a considerable amount of its total storage for tracking them. 
Overall, designing branch predictors by overlooking sparsity leads to suboptimal use of on-chip resources that can greatly affect prediction accuracy. In next section, we introduce a complete architectural design that enables sparsity-aware BP for effectively handling sparsely correlated branches.
%

%
\section{Sparse Predictor Architecture}\label{sec:approach}
\begin{figure}[!t]
\center
\includegraphics[width=0.48\textwidth]{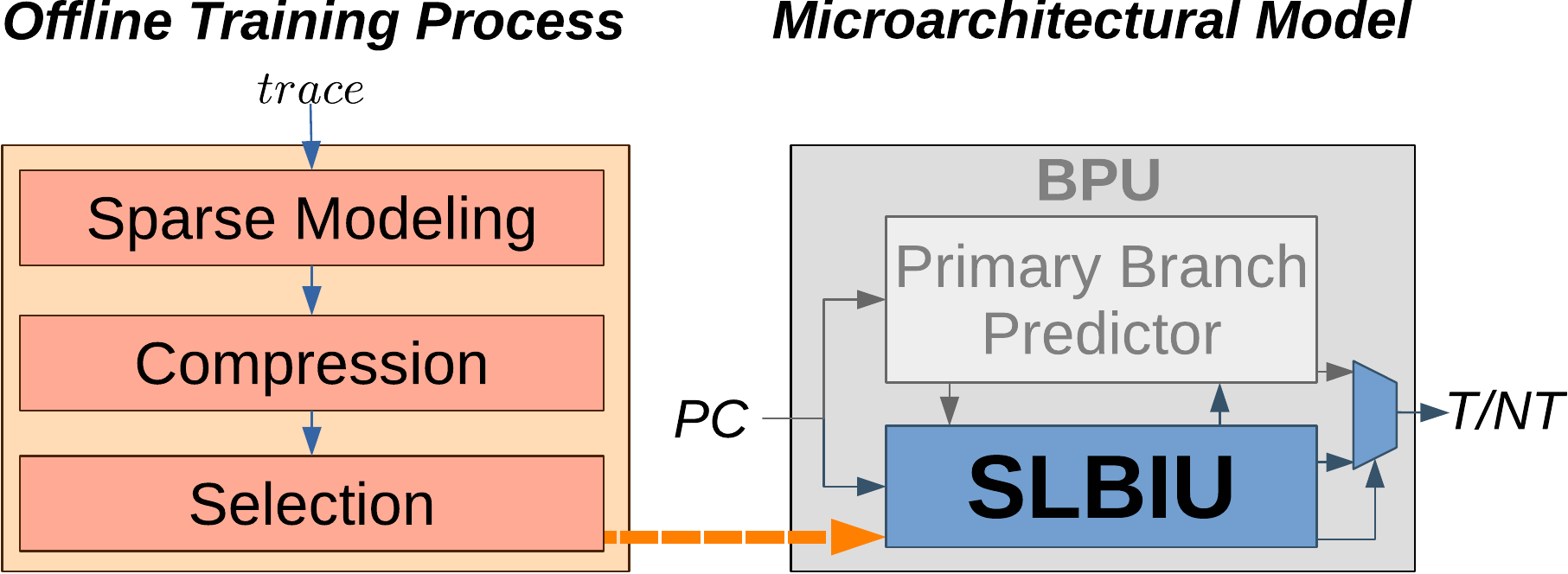}
\caption{Sparsity-aware branch prediction overview.
\label{fig:sparse_predictor_arch}}
\end{figure}

%
We now present our sparsity-aware BP scheme outlined in \cref{fig:sparse_predictor_arch}. Our proposal assumes a deployment scenario where predictions are generated at runtime after an offline training phase, as also considered in the work of Lin \& Tarsa~\cite{bp:not_solved:2019}. Offline training is necessary to capture the predictive statistics of the otherwise hardly detectable sparse correlations. It consists of three major steps, starting with sparse linear modeling for extracting the branch models, then compression of such models and eventually filtering. Branch models are filtered according to certain microarchitectural design constraints that facilitate their use by a dedicated component called {\it Sparse Linear Branch Inference Unit} (SLBIU) for runtime prediction. SLBIU is the BP mechanism that stands at the core of our scheme, specialized to predict branches with offline-prepared sparse models. We envision SLBIU as an auxiliary element of the branch prediction unit (BPU), complementing the functionality of the primary branch predictor that is traditionally trained solely online. In the rest of this section, we describe all the details of our model, including offline processing, microarchitectural modifications and certain system requirements of our deployment scenario.

%
%

%
\subsection{Offline Training Process}\label{sec:approach:offline_process}
%
The offline stage requires a set of branch traces collected by profiling target applications in a post-compilation phase. Such traces undergo \textit{Sparse Modeling} through Lasso logistic regression (as described in \cref{sec:bp_lasso:sparse_modeling}) that produces the \textit{\Lassohints}: a set of per-branch sparse linear models that are attached to the program binary. \Lassohints essentially materialize into a collection of weights and history-indices pairs. Each weight expresses the correlation of the screened branch with the branch at the respective history index. The \lassohints are loaded to the SLBIU using a dedicated SW/HW API and used to perform BP by using the dynamic history as input. The SLBIU functionality and API are described in \cref{sec:approach:slbiu}. As such, it is crucial to keep storage requirements and complexity in reasonable levels without compromising performance. To do so, we perform two necessary optimizations on \lassohints before deployment, denoted as {\it Compression} and {\it Selection} routines in \cref{fig:sparse_predictor_arch}. 
%

%
\subsubsection{Compression}\label{sec:approach:offline_process:compression}
%

%
\textbf{Weights Quantization:} Sparse linear model training is performed in floating point arithmetic with a high-enough precision able to express small parameter updates during models' optimization. However, once sparse models are trained, weights can be represented with fewer bits for use during inference. We quantize the model weights to $Q[I].[F]$ fixed precision format~\cite{gupta2021neural} by rounding each weight to its nearest representable signed number~\cite{quantization:golovin13}. As we will show, 8-bit weights are sufficient for our models expressed in $Q3.4$.
\textbf{History Deduplication:} We observed that, quite often, branch histories lead to (conceptually) duplicated inputs. For example, in a loop-type branch with $s$ iterations, local history satisfies $lhr_{o+s}=lhr_{o+2s}=lhr_{o+3s}=\dots$ for a positive offset $o$. For such a branch, Lasso will assign arbitrary weights in the full history set. However, if there is indeed some correlation with any of these history indices, only one of them is sufficient to express it. To detect these duplicates, we leverage ElasticNet, a generalization of Lasso that assigns approximately the \textit{same} weight on highly correlated or duplicated features~\cite{elasticnet_2005}. In this way, we manage to keep only one instead of multiple identical non-zero weights for branches with such a ``strided'' local-history.
%
\subsubsection{Selection}\label{sec:approach:offline_process:selection}
%
The set of \lassohints produced after compression contains a sparse model for every static branch of the target program. However, evidently, sparse models are not efficient for all static branches. Essentially, the number of static branches per application that are predicted more accurately with a sparse model than a state-of-the-art predictor is conveniently small. In our evaluation, we show that the number of finally selected sparsity hints does not exceed $13$, for a storage overhead of $2$KB. Still, by removing the burden of predicting such sparsely-correlated branches from the primary predictor, enables important improvements, as we discuss in \cref{sec:results}.
As it appears, it is necessary to employ a selection method to filter out the non-promising cases or otherwise to define the cases where it is effective to employ a sparse model. Such selection method needs to solve an optimization problem: identify the subset of branches that are predicted with better accuracy through sparse models under certain storage constraints. We consider two dimensions to express storage constraints, the maximum permitted number of \lassohints, denoted by $\bm{\numoffloadedbranches}$, and the maximum permitted number of non-zero weights per selected hint, denoted by $\bm{nnz}$. Our selection method follows three steps. First, we employ a specific score function $\bm{\score}$ that assigns a scalar score-value to each hint. Positive scores indicate a potential improvement and negative scores a potential drop in performance. Second, the hints with negative scores or with more than $\nnz$ weights are discarded. Finally, the remaining hints are ranked based on their score and the top $\numoffloadedbranches$ (at most) are selected. 
We define two different score functions: \textbf{independent} and \textbf{relative}. In the independent score function, hints with relatively low-accuracy ($<99\%$) are dropped a priori. The remaining hints are assigned with a score equal to the number of their correct predictions solely during the offline analysis. Scores are always positive and they do not consider the performance of the primary predictor. In the relative score function, scores are defined as the difference between the number of correct predictions made by the respective sparse models offline, and the number of correct predictions (estimation with sampling) made by the primary predictor. Both score functions are based on counts of correct predictions instead of accuracy rates to prioritize hints that relate to branches with a greater impact, i.e., hints that achieve high accuracy for branches of very few invocations will have less priority.
%

%
\subsection{Sparse Linear Branch Inference Unit}\label{sec:approach:slbiu}
%
%
\begin{figure*}[!t]
\centering
\includegraphics[width=0.98\textwidth]{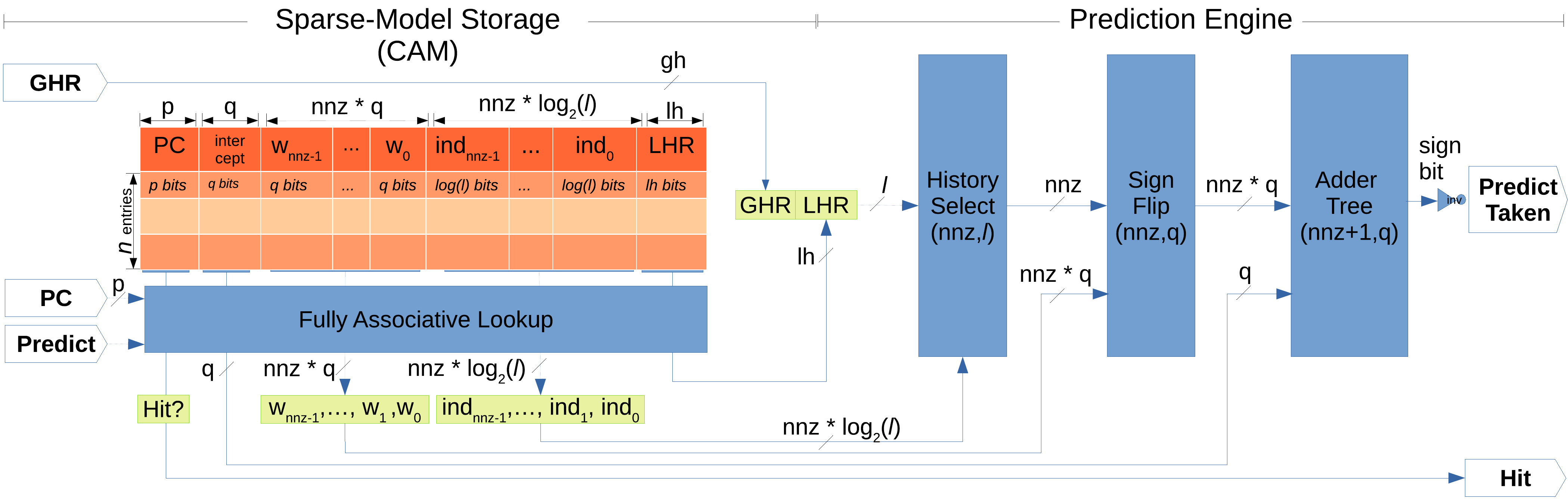}
\caption{\SLBIU architecture with components and functionality. For clarity, the units for loading hints in the CAM space and for dynamically updating LHRs are omitted. The total history bits $l=\numglobalhistbits+\numlocalhistbits$.}
\label{fig:lasso_infer_detailed}
\end{figure*}
%
As mentioned above, the deployment scenario of our scheme involves an offline training phase that produces the sparse prediction models of branches in the form of hints. Sparsity hints are models that receive branch histories as input and provide accurate predictions for the corresponding branches. {\bf SLBIU}, the {\it "Sparse Linear Branch Inference Unit"}, is the hardware mechanism that enables the runtime prediction of branches based on these sparse models which are trained offline. In the rest of this section, we describe the structure and functionality of SLBIU and explain all the microarchitectural modifications required by our sparsity-aware branch prediction scheme.
%
\subsubsection{SLBIU Structure}\label{sec:approach:slbiu:structure}
%
\begin{table}[!t]
\center
\begin{tabular}{cl}
Parameter & Definition \\ \midrule
$\numlocalhistbits$       & Local history length (in bits) \\
$\numglobalhistbits$       & Global history length (in bits) \\
$\numoffloadedbranches$        & Number of branches with sparse models \\
$\nnz$      & Maximum number of non-zero weights  \\
$\numweightbits$        & Weights bit-width \\
$\numpcbits$        & Branch PC bit-width \\ \bottomrule       
\end{tabular}
\caption{SLBIU architecture knobs.}
\label{tab:slbpu_params}
\end{table}
\cref{fig:lasso_infer_detailed} depicts in detail the building blocks of SLBIU, along with all their essential design parameters described in \cref{tab:slbpu_params}. SLBIU is abstractly divided in two main parts, the {\it Sparse-Model Storage}, that is a content-addressable memory (CAM) space, and the {\it Prediction Engine}, that is an arithmetic logic circuitry. The CAM space is fully associative PC-based and holds the \lassohints and the local histories of the (static) branches selected during the offline training process. Recall that \lassohints represent sparse models. We use the Coordinate storage format (COO) to encode the sparsity hints, for keeping a fixed size per hint-vector (zero padding) and for allowing a convenient hardware implementation. 
SLBIU can store up to $\numoffloadedbranches$ hint-vectors in its CAM space. Each hint-vector includes the $\numweightbits$-bit wide $\nnz$ weights and their respective history indices. Indices are $\lceil\log_2(\numglobalhistbits+\numlocalhistbits)\rceil$ bits wide as they encode a history position in the range $[0,\numglobalhistbits+\numlocalhistbits-1]$. The intercept values (one per sparse model) and the weights are reduced to $\numweightbits$ bits after the quantization step discussed in \cref{sec:approach:offline_process:compression}. Eventually, considering all the design parameters listed in \cref{tab:slbpu_params} ($\numlocalhistbits,\numglobalhistbits,\numoffloadedbranches, \nnz, \numweightbits, \numpcbits$), the amount of storage space required by SLBIU is defined as:
%
\begin{equation}\label{eqn:slbpu_storage}
    \text{Storage} \triangleq \numoffloadedbranches \left(\numpcbits + \numweightbits + \nnz  \numweightbits + \nnz \lceil\log_2(\numlocalhistbits+\numglobalhistbits)\rceil + \numlocalhistbits \right){\scriptsize.}
\end{equation}
%
As it appears, the SLBIU storage scales logarithmically with the global history length ($\numglobalhistbits$). Considering that current branch predictors already employ global histories of the order of thousands previous branch outcomes (bits), such logarithmic relation essentially discharges the history length from typically being the major limiting factor. On the other hand, the linear relation of the required storage with the number of branches with sparse models ($\numoffloadedbranches$) and with the number of their non-zero weights ($\nnz$), makes both $\numoffloadedbranches$ and $\nnz$ the most crucial parameters. In \cref{sec:results} we discuss the trade-offs that occur from the non-trivial task of determining the optimal values of $\numoffloadedbranches$ and $\nnz$ under certain storage budgets.
%
\subsubsection{SLBIU Functionality}\label{sec:approach:slbiu:functionality}
%
As outlined by \cref{fig:sparse_predictor_arch}, in our design, SLBIU does not represent a standalone branch predictor. Essentially, the utility of SLBIU is revealed when coupled with the primary branch predictor of a CPU design through the improvement of the prediction of sparsely correlated branches. In our deployment scenario, SLBIU is informed of the branches with accurate sparse models after an offline analysis. That is, we assume that SLBIU is initialized before a program's execution phase starts. Initialization includes loading all the sparse model parameters of the selected branches trained offline (intercept value, $nnz$ weights and history indices, as shown in \cref{fig:lasso_infer_detailed}) and resetting all the respective LHR fields to zero. Nevertheless, the potential of performing sparse modeling purely online is not generally excluded. In \cref{sec:online_lasso} we briefly discuss the feasibility of such an implementation scenario, although we keep it out of the scope of this work.
Eventually, SLBIU contains one sparse model per each static branch that has been selected at the end of the offline training phase. When branches are routed for prediction in BPU, SLBIU is probed based on the branch PC in parallel with the primary branch predictor. When discovering that a branch possesses an entry in SLBIU (\textit{"hit"}), the main branch predictor is signaled to halt any update of its internal state related to that branch, i.e., entries allocations and/or update of entries state machines. Thereafter, the prediction of that branch happens exclusively by SLBIU. In that sense, the selected sparsely correlated branches are \textit{offloaded} to SLBIU, since they do not require resources by the primary predictor anymore. However, any branch history organization that is used and maintained in BPU is still normally updated.
In particular, we assume that BPU manages the global branch history with a single GHR which is common for the primary predictor and SLBIU. The fraction of GHR that each component uses internally for prediction and update can be different, depending on the mechanism. In our evaluation we couple SLBIU with \tagesclSmall that uses GHR slices of up to $1,000$ bits long, whereas in SLBIU we have found that a $512$-long global history suffices to effectively capture the potential for improvement in our dataset. Below we describe the prediction and update process in SLBIU.
\textbf{Prediction computation:} After initialization, GHR along with the per-branch maintained LHR are the two main inputs of SLBIU for predicting an offloaded branch. At prediction time, LHR and GHR are concatenated forming an $l$-bit long history vector from which the important $\nnz$ bits are selected through an array of $\nnz$ $l$:1-multiplexers (\HistorySelect). Predictions are based on the dot product of the important $\nnz$ history bits and their $\nnz$ weights. An adder tree can be used to sum up the vector of products and the intercept value. Similarly to previous work~\cite{perceptron2001}, as we interpret \textit{taken/not-taken} events with $\{-1,+1\}$ values (usually expressed with $\{0,1\}$ in GHR/LHR), respectively, vector multiplication happens practically by only flipping the weights' signs that are paired with \textit{not-taken} history bits (\SignFlip). The prediction of a branch is \textit{not-taken} if the dot-product sign is negative, and \textit{taken} otherwise. To allow higher clock rates and to limit the latency to acceptable 3-4 cycles \cite{zangeneh2020branchnet, zhao2021cobra}, we pipeline the prediction operation of \SLBIU into $3$ stages: 1\textsuperscript{st} stage performs the fully associative lookup, 2\textsuperscript{nd} stage extracts the model weights and LHR from the CAM, concatenates and selects history bits and flips weights signs, while 3\textsuperscript{rd} stage is dedicated to the adder-tree.
\textbf{Update:} The sparse models of the branches offloaded to SLBIU are only updated once at initialization. That is, at a context switch or at the start of a program phase for which a different set of hints has been produced offline. Our design assumes that a different set of \lassohints can be loaded in SLBIU per program phase. \cref{sec:approach:slbiu:sys_requirements} clarifies the requirements of such an approach. During execution, only LHRs are updated according to the respective branch outcomes. To support a simultaneous update of LHRs alongside the retrieval of model parameters for prediction, we implement a dual-ported CAM storage space, i.e., a single-bit port for writing and an entry-wide port for reading.
%
\subsection{SW/HW Interface for Sparsity Hints}\label{sec:approach:slbiu:sys_requirements}
%
In this paper, we advocate the use of offline training for more accurate and focused sparse modeling. Our scheme relies on a sparsity analysis performed at compilation over traces that contain branch outcomes recorded from normally running the application. The traces can be generated through a profile-guided optimization (PGO) phase~\cite{Gupta02profileguided}; through manual workload optimization; or through a JIT analysis~\cite{JIT_tracing}. Recent work also suggests that PGO may be performed over a wide corpus of other applications using ML techniques~\cite{PGO_ML}.
In their previous work~\cite{bp:not_solved:2019}, Lin and Tarsa indicate that acquiring several traces from a single program allows to effectively refine training over specialized programs statistics. Therefore, we obtain multiple traces representing different phases of a program's execution (using SimPoint~\cite{hamerly2005simpoint} intervals) resulting in a respective set of trained sparse models. These models pass the selection and compression steps individually so they may each focus on a different subset of branches that dominate each specific program phase. Even if a branch appears in different models, its weights may differ, representing its localized behavior.
The sparse models are stored in the binary as program metadata. According to our experiments (as seen on Section~\ref{sec:results:sensitivity}), the binary size overhead is expected to be minimal, since, per execution phase, sparse models that account for less than $2$KB of storage space suffice to capture effectively the majority of mispredictions from sparse branches. For the on-chip delivery of model parameters, the binary can be annotated with trigger points that will indicate the passing of program phases. The application  will then be responsible for unpacking the sparse model of each phase to memory and, using dedicated instructions (ISA extensions), to populate the SLBIU weights from that memory range. Given that the storage overhead of each model is small and the phases represented are multiple billions of instructions each (depending on simpoint representativeness), the overall loading time is also assumed to be negligible. We leave a more detailed analysis and evaluation of the requirements of such a SW/HW interface for sparsity hints or other alternative approaches to future work.

%
\section{Evaluation Methodology}\label{sec:evaluation_methodology}
%
%
\subsection{Experimental Setup \& Benchmarks}\label{sec:evaluation_methodology:setup}
%
We have implemented the full process of offline training as a software module that receives branch traces and produces the set of selected branch sparse models. \textit{Sparse modeling} is performed through Lasso logistic regression. 
Training on branches with roughly $10$ million dynamic executions and $1,024$-bit long branch history features takes around $3-4$ minutes in a commodity server.
To evaluate our sparsity-aware BP scheme, we implemented SLBIU and the interface for initialization with the offline-produced sparse models in the \cbpfive trace-based framework~\cite{cbp2016}. TAGE-SC-L predictors are configured to ignore the branches that \textit{hit} in SLBIU. TAGE-SC-L and SLBIU work in tandem based on a common GHR. We also model the program-phase adaptation of sparse models by initializing SLBIU before each trace simulation, assuming that each trace represents a program phase with different sparse models. As we perform simulations in \cbpfive framework, we evaluate our BP model exclusively based on MPKI measurements. In this way, we are able to gauge the improvement of state-of-the-art BP independently from the various design-specific artifacts of a modern CPU. Nonetheless, this paper includes all the necessary details for a full-scale evaluation. 
We evaluate our microarchitectural model over a rich set of traces that undergo sparse modeling before simulation. Our trace pool includes the publicly available set of \cbpfive~\cite{cbp2016} traces and a set of traces recorded from running the $10$ \specint Rate benchmarks~\cite{spec2017} using the \textit{ref} input-set. SPEC traces are obtained for each of the $100$M-instructions long Simpoints that we identify per benchmark ($20$ on average). Therefore, in our experiments, SLBIU is reconfigured every $100$M instructions for \specint and every $400$M instructions on average on CBP-5 traces, i.e., once per trace. In total, we use $382$ \cbpfive traces, after dropping some duplicated and corrupted traces and also a few that are already highly optimized by \tagesclSmall ($<0.01$ MPKI). For SPEC benchmarks with traces from several inputs (\emph{xz}, \textit{x264}, \textit{perlbench}, \textit{gcc}) we report average metrics with ranges. Note that all our measurements concern the total trace simulation, i.e., no warm-up phase exists. As we are interested in cases where storage pressure is high, we couple SLBIU with \tagesclSmall and we report the relative difference in the obtained MPKI. As such, we (arguably) emulate scenarios where large predictors are cornered by applications with extensive working sets. By lacking such cases in our dataset, we do not observe significant differences when coupling SLBIU with \tagesclBig. Therefore, we do not present quantitative analysis for the $64$KB variant of TAGE-SC-L. Yet, the 8KB configuration of TAGE-SC-L that we study resembles closely a practical BP design for today's common CPU resource budgets, as also considered by previous work~\cite{bp:not_solved:2019}.
\vspace{5mm}
%
%
\subsection{Physical Implementation}\label{sec:physical implementation}
%

We implemented \SLBIU as presented in Section~\ref{sec:approach:slbiu} with 
 an industry-grade technology library and compiler tools. The place-and-route was performed at typical corner (0.8\,V, 25C). We use retiming to balance the 3 pipeline stages, automatic clock gating, and manual clock gating based on the \textit{hit} signal to avoid switching in the history select and compute logic. The CAM is a fully-associative lookup table implemented as a register file of $\numoffloadedbranches$ entries of $\numpcbits+\numweightbits+\nnz\cdot(\numweightbits+ \lceil\text{log}_2(\numlocalhistbits+\numglobalhistbits)\rceil+\numlocalhistbits)$ bits each. 
The power is evaluated based on the switching activities (VCD) extracted from timing-annotated (SDF) post-place \& route gate-level simulations running a synthetic benchmark sweeping over different scenarios of branch ratio and branch offloading rate. A scenario consists of a randomly generated \textit{trace} and a set of random \textit{\lassohints}. The scenarios are parameterized by ``branch frequency'' (ratio between the number of dynamic branches and the number of all dynamic instructions), and by  ``offloaded branch ratio'' (ratio between the number of offloaded \lassohints and the total number of static branches in the trace). Instructions are scheduled uniformly across the trace, while hints' weights and indices are drawn from a uniform distribution. Using synthetic scenarios allows us to explore various extreme cases of read/write intensity to our proposed \SLBIU circuit.
The simulation of \SLBIU is embedded in a \texttt{cocotb}-based \cite{rosser2018cocotb} testbench, which begins with the \lassohints initialization (the entire CAM is always filled with hints) and followed by the execution of the 10K-long trace, fetching an instruction every clock cycle. For each branch instruction of the trace the corresponding inputs are applied to \SLBIU (PC, GHR and Predict signals), the prediction output is compared with the expected result to verify correctness, and the LHR is updated based on branch resolution.
%
\section{Results \& Analysis}\label{sec:results}
%

We start our analysis by quantifying the effects of various design parameters, then, we continue with an overall evaluation of our approach, and finally, we demonstrate the effectiveness we observe at the circuit level. In all our experiments, \tagesclSmall, as implemented in \cbpfive~\cite{cbp2016}, is the primary branch predictor coupled with the respective SLBIU configuration that is examined. All evaluations are relative to the standalone \tagesclSmall. Overall, our results demonstrate that our design can achieve noticeable MPKI improvements with insignificant storage overheads.
%

%
\subsection{Sensitivity to Design Parameters}\label{sec:results:sensitivity}
%
%
\begin{figure}[t]
\centering
  \begin{subfigure}[t]{1.0\linewidth}
    \includegraphics[width=1.0\linewidth]{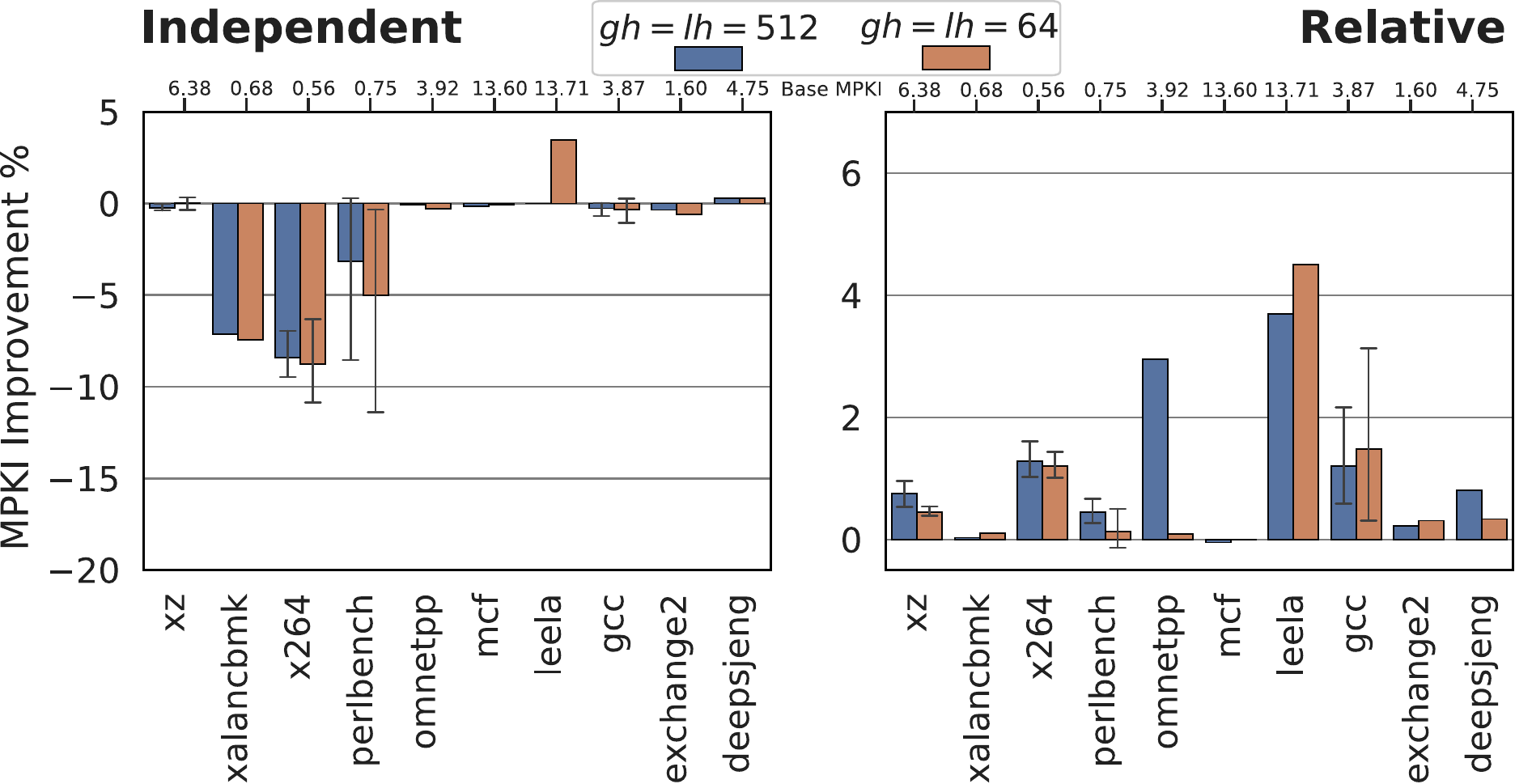}
    \caption{SLBIU 0.5\,KB}
    \label{fig:results:quantifying_hist_adaptivity_size_05kb}
  \end{subfigure}
  
  \begin{subfigure}[t]{1.0\linewidth}
    \includegraphics[width=1.0\linewidth]{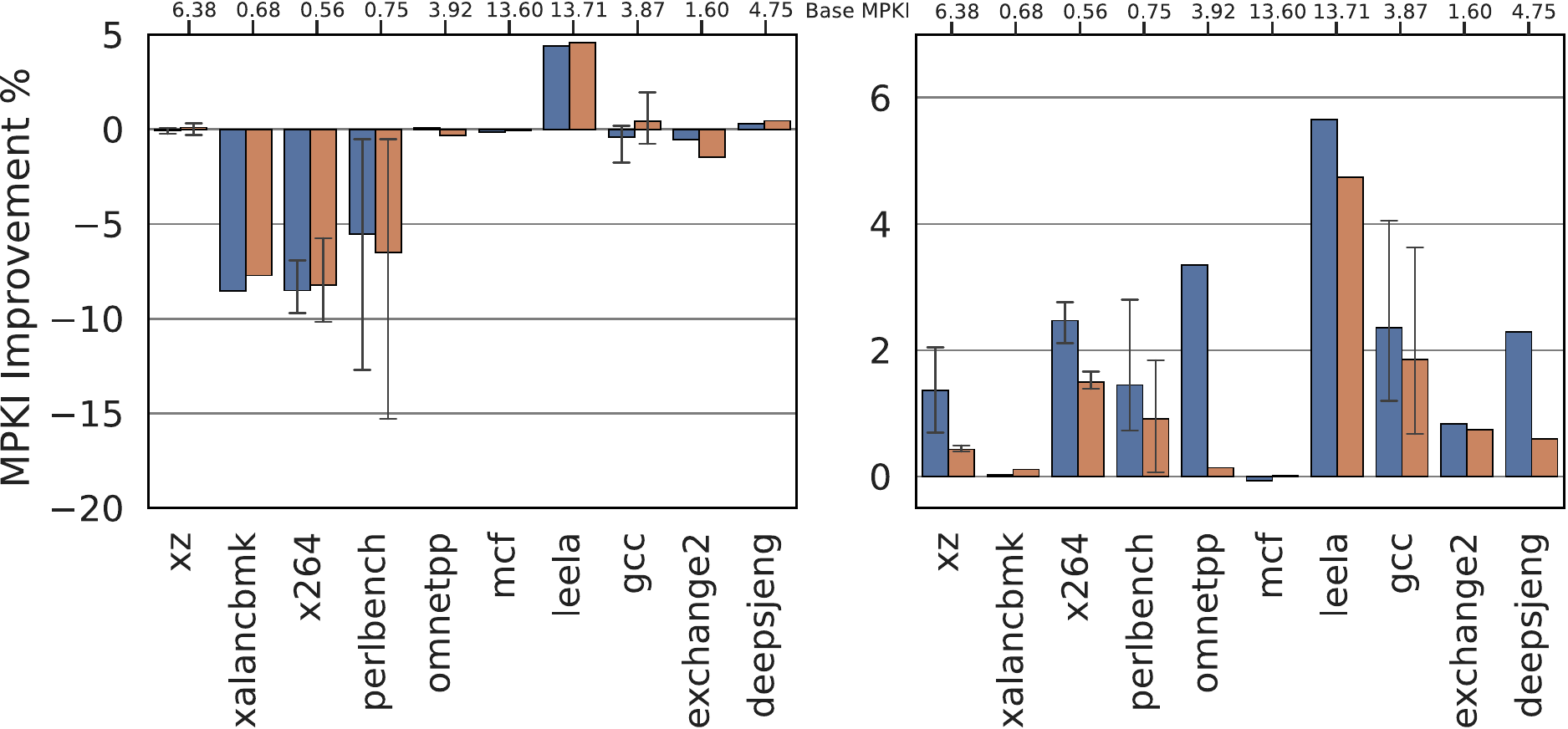}
    \caption{SLBIU 2\,KB}
    \label{fig:results:quantifying_hist_adaptivity_size_2kb}
  \end{subfigure}

  \begin{subfigure}[t]{1.0\linewidth}
    \includegraphics[width=1.0\linewidth]{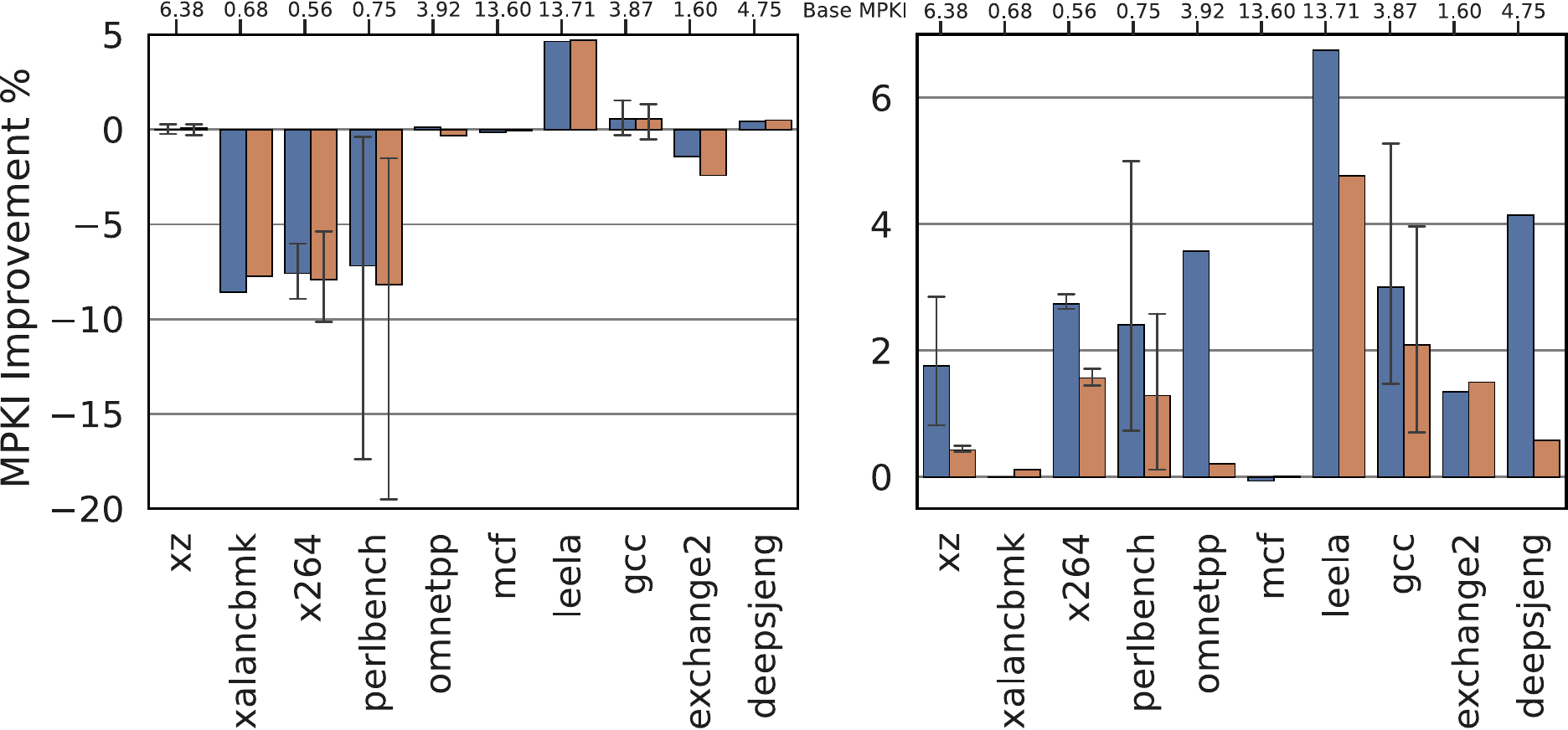}
    \caption{SLBIU 8\,KB}
    \label{fig:results:quantifying_hist_adaptivity_size_8kb}
  \end{subfigure}
  \caption{\label{fig:results:sensitivity} Impact of design parameters (higher is better). The MPKI of baseline \tagesclSmall is on top $x$-axis.}
\end{figure}
%
We distinguish $4$ main factors that can affect the performance of our design: the amount of available storage, the method of hints selection, the length of branch histories and the quantization degree of models' weights. As these design parameters are interdependent, to identify their impact, we define a specific range of cases to experiment, and then, we perform MPKI evaluation by simulating our SPEC traces. Eventually, after experimenting with several combinations, we analyze the most interesting and representative cases.
In particular, we choose to experiment with $3$ different storage budgets, the smallest being $0.5$KB, a moderate option equal to $2$KB, and the largest at $8$KB. We perform \textit{Sparse Modeling} (as described in \cref{sec:evaluation_methodology:setup}) and we prepare a set of branch sparse models for two different history sizes, $\numlocalhistbits=\numglobalhistbits=64$ and $\numlocalhistbits=\numglobalhistbits=512$. In our first set of experiments, to fully capture the potential of branch sparse models, we allow full-precision floating point weights ($32$-bit wide) without quantization. Thereafter, we run our two methods for hints selection, \textit{independent} and \textit{relative}, to identify the optimal sets of sparsity hints that satisfy the three storage budgets examined. In practice, each selected set defines the actual dimensions $\numoffloadedbranches,\nnz$ of SLBIU's CAM space. Eventually, the $\numoffloadedbranches,\nnz$ pair we choose is the one with the highest sum of scores. We configure SLBIU according to the chosen $\numoffloadedbranches,\nnz$ pair and we simulate our SPEC trace-set.
\cref{fig:results:sensitivity} depicts the results of our analysis in two distinct columns, left side for \textit{independent} and right side for \textit{relative} selection. Horizontal figure-pairs represent the three storage budgets we examine comparing the two different history sizes used in SLBIU. Unsurprisingly, the \textit{relative} selection consistently outperforms the \textit{independent} selection by being able to prioritize only the hints that are guaranteed to improve \tagescl. Noticeable improvements can be seen in almost all the benchmarks, except from \texttt{mcf}, with \texttt{leela} demonstrating the largest uplift in all cases. As expected, the benefit increases for larger storage budgets where more sparse models can be employed, as validated by \cref{tab:results:n_nnz_fp} listing the chosen $\numoffloadedbranches,\nnz$ pairs. The maximum amount of offloaded branches increases almost linearly with the available storage, while sparsity remains steadily high. The $8$KB SLBIU configuration features the maximum $nnz$, that of $43$, whereas for both the other storage configurations, $nnz$ does not exceed $34$.
%
\begin{table}[ht]
\centering
\footnotesize
    \begin{tabular}{lccccc}
    \toprule
                 & \multicolumn{2}{c}{Independent} &       & \multicolumn{2}{c}{Relative} \\
\cmidrule{2-3}\cmidrule{5-6}      & \emph{\numlocalhistbits=\numglobalhistbits}=64    & \emph{\numlocalhistbits=\numglobalhistbits}=512   &       & \emph{\numlocalhistbits=\numglobalhistbits}=64    & \emph{\numlocalhistbits=\numglobalhistbits}=512 \\
    0.5KB        & (5,16) & (3,17) &       & (5,16) & (2,34) \\
    2KB          & (18,19) & (8,34) &       & (13,28) & (8,34) \\
    8KB         & (50,29) & (29,39) &       & (46,32) & (27,43) \\
    \bottomrule 
    \end{tabular}%
\caption{Chosen pairs of ($\numoffloadedbranches,\nnz$) for \cref{fig:results:sensitivity}.}\label{tab:results:n_nnz_fp}
\end{table}
%

%
Furthermore, the set of $512$-bit histories account for higher improvements, with a few exceptions in \texttt{gcc} and \texttt{leela} at $0.5$KB, and \texttt{exchange2} at $8$KB, where $64$-bit histories perform better. Therefore, larger histories broaden the scope of sparse models and allow them to capture effectively correlations that are found in the quite distant past. Even more so, as illustrated in \cref{tab:results:n_nnz_fp}, they achieve that by requiring storage for only the quite few important segments of the history. As our experiments dictate, in the rest of our analysis we will focus on sparsity hints over $\numlocalhistbits=\numglobalhistbits=512$-bit histories, filtered with the \textit{relative} selection method.
%
%
\begin{figure}[t]
\centering
  \begin{subfigure}[t]{0.475\linewidth}
    \includegraphics[width=1.0\linewidth]{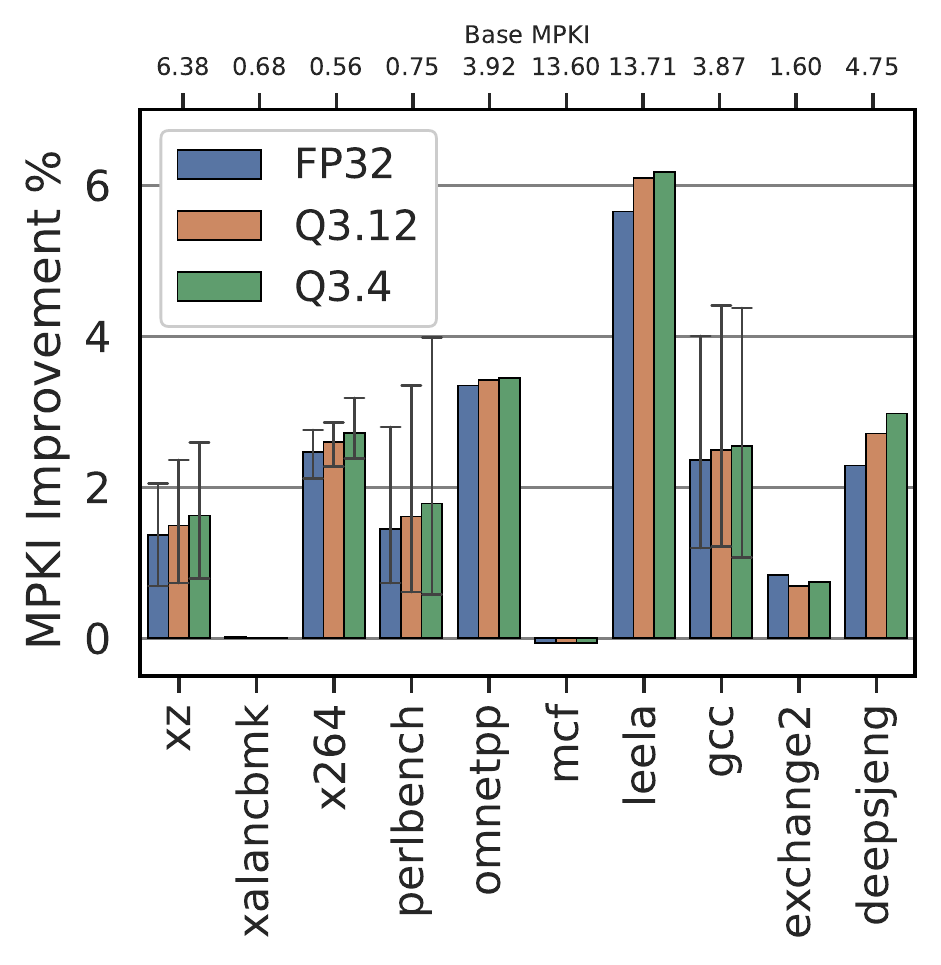}
    \caption{SLBIU $2$KB}
    \label{fig:results:quantifying_quantization_2kb}
  \end{subfigure}
  \begin{subfigure}[t]{0.475\linewidth}
    \includegraphics[width=1.0\linewidth]{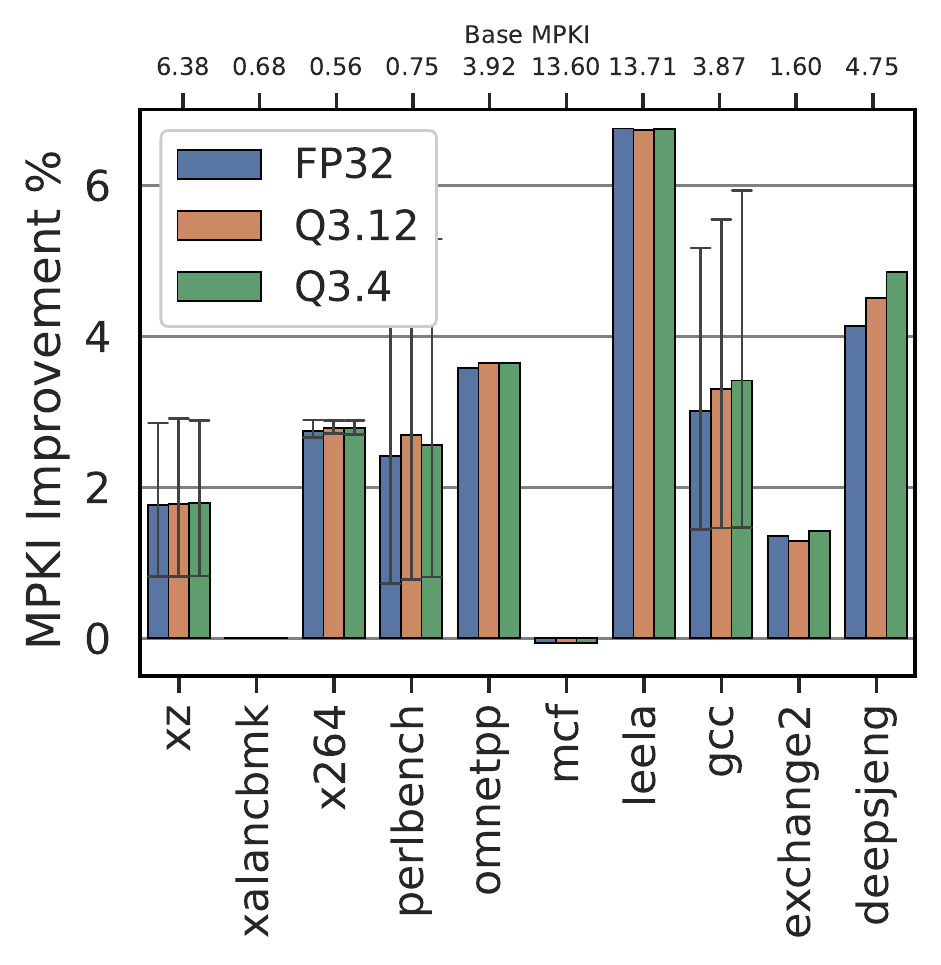}
    \caption{SLBIU $8$KB}
    \label{fig:results:quantifying_quantization_8kb}
  \end{subfigure}
  \caption{\label{fig:results:quantization} Effect of hints quantization (higher is better). The MPKI of baseline \tagesclSmall is on top $x$-axis.}
\end{figure}
%
%
Next, we explore the impact of quantization, the $4^{th}$ important performance factor of our design. We evaluate the effectiveness of our model with $8$- and $16$-bit quantization degrees using $Q3.4$ and $Q3.12$ signed fixed-point formats, respectively. To do so, we simulate our SPEC traces using the two most promising SLBIU configurations of $2$KB and $8$KB. In \cref{fig:results:quantization} we plot the MPKI improvements obtained with quantized models, comparing them with full-precision. 
%
\begin{table}[ht]
  \centering
  
    \begin{tabular}{cccc}
    \toprule
          & FP32  & Q3.12 & Q3.4 \\
    \midrule
    2KB   & (8,34) & (11,34) & (13,36) \\
    
    8KB   & (27,43) & (33,53) & (53,42) \\
    \bottomrule
    \end{tabular}%
  \caption{Chosen pairs of ($\numoffloadedbranches,\nnz$) for \cref{fig:results:quantization}}\label{tab:results:n_nnz_quantization}
\end{table}%
%
According to our results, MPKI improvements are successfully sustained after quantization, although no significant gains are observed. More specifically, for the $2$KB SLBIU configuration, the benefits are mostly higher at $Q3.4$ ($8$-bit) than at $Q3.12$ ($16$-bit) format, manifesting the available quantization headroom. With $8$KB SLBIU, such trend is observed only in \texttt{gcc} and \texttt{deepsjeng}. In most benchmarks, improvements tend to saturate along all precision formats, with the exception of \texttt{perlbench}, where the $Q3.12$ format performs marginally better. Naturally, quantization enables storage savings that can allow more branches to be offloaded to \SLBIU. Our evaluation reveals that such opportunity is more important in lower storage budgets. That is, quantization appears to be necessary for minimizing storage requirements without compromising prediction accuracy.
%
%
\begin{figure*}[t]
\centering
\includegraphics[width=0.98\linewidth]{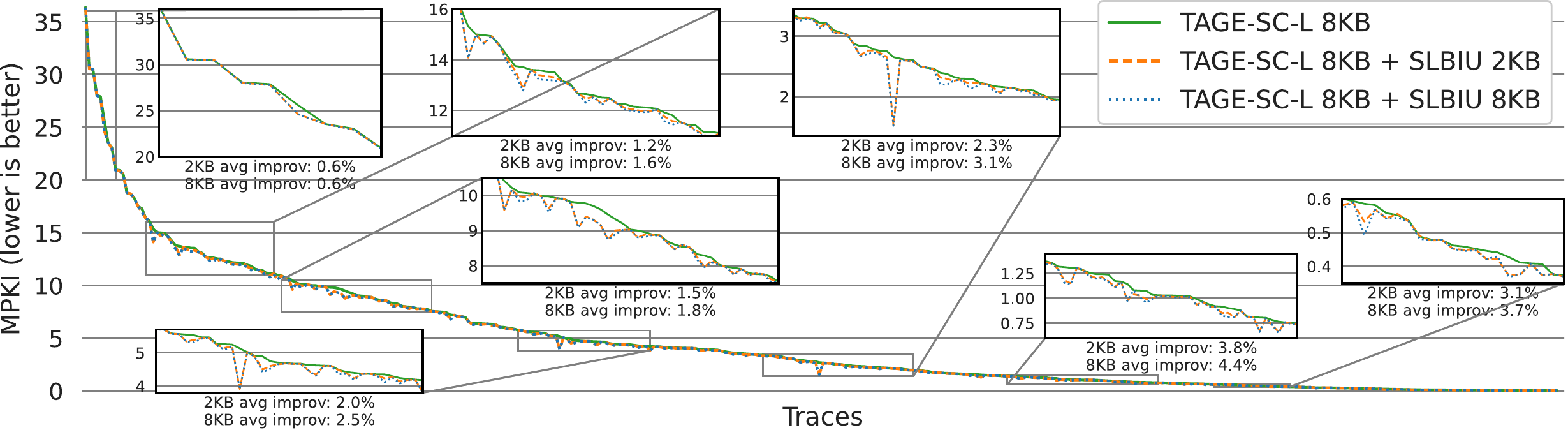}
  \caption{Large-scale MPKI comparison with 2KB and 8KB of SLBIU storage, sorted based on \tagesclSmall
  }
  \label{fig:results:sea_of_traces}
\end{figure*}
%

%
In \cref{tab:results:n_nnz_quantization} we present the $\numoffloadedbranches,\nnz$ pairs chosen in our experiment, confirming the increase of $\numoffloadedbranches$ from full-precision to quantized models. More importantly, \cref{tab:results:n_nnz_quantization} shows that sparsity levels remain comparable; $nnz$ is equal to $36$ and $42$ for the configurations of $2$KB and $8$KB, respectively. That is, despite the $4\times{}$ relation of available storage, offloaded branches satisfy well a similar sparsity threshold. Recall that $nnz$ specifies the maximum number of weights per offloaded branch by defining the width of SLBIU's CAM (see \cref{fig:lasso_infer_detailed}). As such, configuring SLBIU efficiently for exploiting most of the underlying opportunity appears to be feasible. In next section, we explore this aspect by evaluating the effectiveness of specific SLBIU designs over our large set of traces.
%

%
\subsection{Large-scale MPKI Evaluation}\label{sec:results:large_scale_mpki_evaluation}
%

\begin{figure}[t]
\centering
\includegraphics[width=0.48\textwidth]{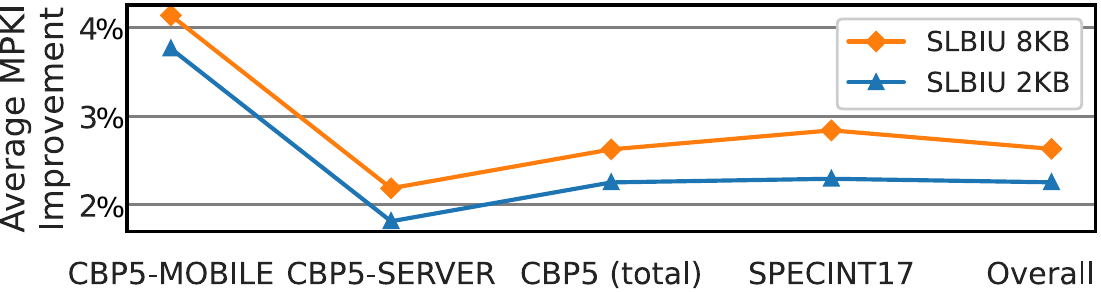}
\caption{Average MPKI improvements compared to the standalone \tagesclSmall for different trace groups.}
\label{fig:results:summary:average}
\end{figure}

We evaluate the performance of the two best performing configurations from our previous experiments over our full trace set (described in \cref{sec:evaluation_methodology:setup}). In particular, we compare SLBIU of $2$KB and $8$KB for $512$-bit histories, where sparse-models' weights are quantized to $8$ bits ($Q3.4$ format) and the set of offloaded branches is selected with the \textit{relative} method, satisfying the $\numoffloadedbranches,\nnz$ dimensions in \cref{tab:results:n_nnz_quantization}, i.e., there can be up to $\bm{\numoffloadedbranches}$ selected branches per execution phase (trace) with less than $\bm{nnz}$ weights.  
\cref{fig:results:sea_of_traces} depicts an S-curve of MPKI for the 392 traces of our evaluation set (\texttt{x}-axis) sorted according to \tagesclSmall MPKI. 
We refer to the improvements of the 2KB/8KB configurations in 3 MPKI ranges. In the high range of above 5 MPKI (128 traces) our configurations reduce on average 0.13/0.15 MPKI (1.3\%/1.6\%). In the middle range of 1-5 MPKI (141 traces) our configurations reduce on average 0.05/0.07 MPKI (2.1\%/2.7\%). In the low range of 0.01-1 MPKI (123 traces) our configurations reduce on average 0.012/0.014 MPKI (3.4\%/3.7\%). Note, that across various MPKI ranges we also observed $47$ traces (for both designs) where no branches were offloaded to SLBIU resulting in no fluctuation in MPKI. Interestingly, in $23$ traces the $2$KB configuration achieves (marginally) the best performance. Essentially, such phenomenon exposes the importance of the selection method in large storage budgets, that needs to be optimized adequately for balancing storage exploitation and performance. It also shows that the $2$KB SLBIU configuration can be a highly effective design.
\cref{fig:results:summary:average} compares the mean MPKI improvements of the two SLBIU configurations for different groups of traces. Although the storage budget of 8KB gives a higher improvement across all groups of traces, it is only $0.38\%$ higher than of the $2$KB. This demonstrates that large storage budgets are not necessary to capture effectively the underlying opportunity.
%

%
%

%
\subsection{Circuit-level Evaluation}\label{sec:results:hweval}
%

%
\begin{figure}[t]
  \begin{subfigure}[t]{1.0\linewidth}
    \includegraphics[width=1.0\linewidth]{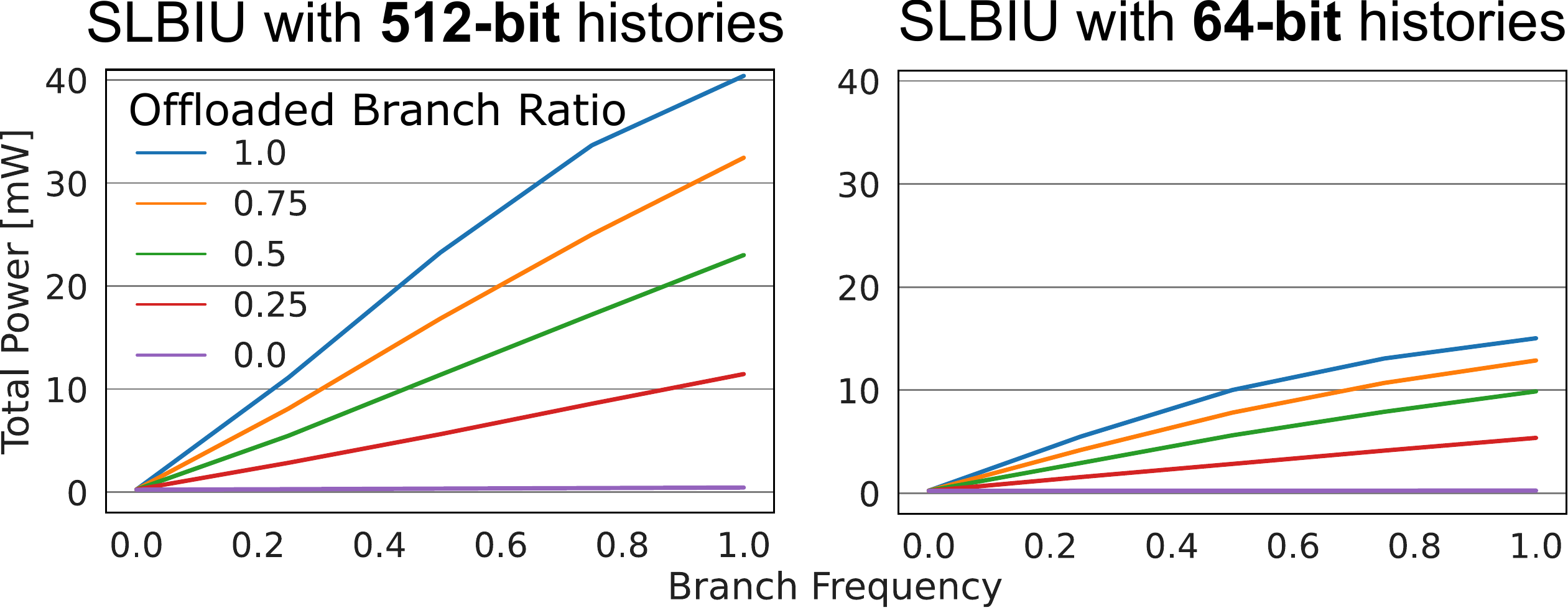}
    \caption{Total power of SLBIU. \label{fig:rtl:module_total_power}}
  \end{subfigure}

  \begin{subfigure}[t]{1.0\linewidth}
    \includegraphics[width=1.0\linewidth]{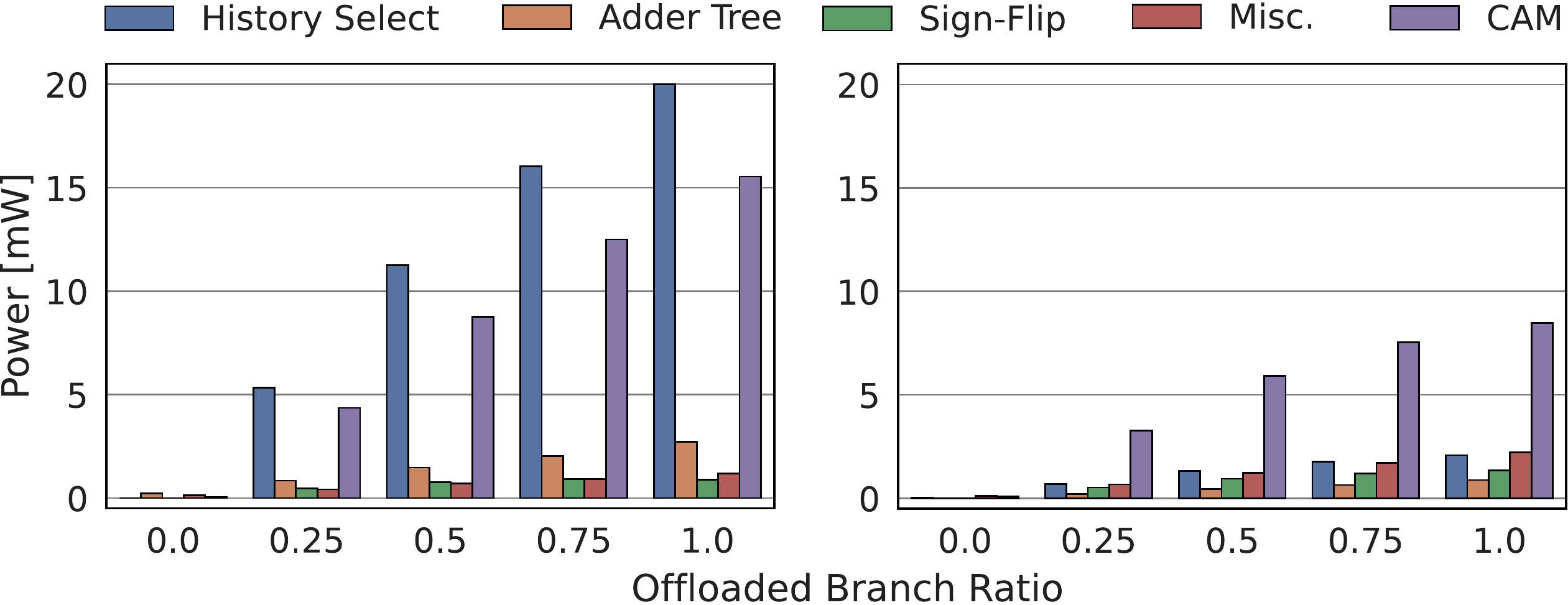}
    \caption{Power of \SLBIU's components, at $100\%$ branch frequency.\label{fig:rtl:component_dynamic_power}}
  \end{subfigure}
\caption{Power consumption of two \SLBIU configurations on synthetic scenarios.}
\end{figure}
%
%

%
We evaluate the two \SLBIU configurations $(\numlocalhistbits=\numglobalhistbits=512,\numoffloadedbranches=13,\nnz=36,\numweightbits=8)$ and $(\numlocalhistbits=\numglobalhistbits=64,\numoffloadedbranches=20,\nnz=29,\numweightbits=16)$ that achieved the best MPKI improvement on \specint under a $2$KB storage budget, for the long and the short history lengths, respectively. The evaluation is in terms of \emph{timing}, \emph{area} and \emph{power} with respect to 28\,nm technology. We also provide a rough estimate of this evaluation with respect to a 7\,nm technology, for which we use the following scaling factors from 28\,nm to 7\,nm: 0.4x power@same speed, 1.8x speed@same power, 3.4x area\cite{techscaling,7nm}.
\textbf{Timing.} Both designs run at 750/790\,MHz in 28nm. (1.4 \,GHz in 7\,nm). The 3 pipeline stages are balanced, with the critical path dictated by the adder tree (3\textsuperscript{rd} stage) due to the 16-bit operands in the short history configuration, and by the history selection (2\textsuperscript{nd} stage) in the long history configuration. Note that in 7nm same 3-cycle latency can also be retained under certain frequency requirements~\cite{zhao2021cobra,zangeneh2020branchnet}.
\textbf{Area.} 
For both designs, the standard cell-based content-addressable memory (CAM) is dimensioned to 2\,KB and thus, it dominates the area breakdown (98\%/88\%). Nonetheless, CAM occupies just 0.34\,mm$^2$ of area (0.1\,mm$^2$ in 7nm).
\textbf{Power.} \cref{fig:rtl:module_total_power} shows the power for both modules for various offloading ratios and branch frequencies. At full utilization (100\%/100\%), the two candidate modules consume 15 or 40\,mW (i.e., up to 28\,mW at 1.3\,GHz in 7\,nm), where at zero utilization requires just 220 or 240\,\textmu W thanks to the high switching reduction achieved by our manual clock gating scheme. Importantly, the \SLBIU spends only negligible power for the lookups in its CAM that result in a miss (purple line in~\cref{fig:rtl:module_total_power}). Conversely, the power is effectively spent only in the cases of a hit. \cref{fig:rtl:component_dynamic_power} shows a breakdown of power consumption of each component of \SLBIU (CAM, Adder Tree, etc.) at the most aggressive, yet realistic scenario with 100\% branch frequency (the unit queried every clock cycle) and over varying offloaded branch ratios. Most power is spent in the history-select unit (50\%/13\%) and the CAM register file (39\%/56\%). Despite the same storage, the power consumption is smaller for the 64-bit history configuration due to the 16$\times{}$ smaller history select unit and narrower CAM.
%

%
\section{Online Sparse Modeling}\label{sec:online_lasso}
%
In our study, we have shown that sparse correlations of branches with branch history can be detected efficiently offline with sparse modeling. In this section, we briefly argue that such sparsity can be detected also with online training.
To that end, we implemented sparse linear modeling in an online setting, where model parameters are updated after predictions are resolved during trace simulation. We employ $512$-bit long global and local histories for comparing online with offline findings from \cref{tab:motivating_examples}. We experimented with several optimization methods and we found stochastic gradient descent with cumulative penalty (SGD-L1)~\cite{lasso:sgd_cum_l1} as the most efficient. We improved SGD-L1 by adapting the hyperparameter $\lambda$ with online binary search, i.e., starting with $\lambda=0.01$ and halving it or doubling it within the range $[1e-5,0.1]$ to keep the number of non-zero model weights at most $50$. 
%
\newcommand{\spspa}[0]{\hspace{-3mm}}
\begin{table}[ht]
\footnotesize
\begin{center}
  \begin{tabular}{lrrr}
	\toprule

     {\footnotesize Trace / PC}	&  \multicolumn{2}{c}{\footnotesize \texttt{Online} (nnz)} & {\footnotesize  \texttt{Offline}} (nnz) \\
     \midrule
     {\scriptsize LONG-MOBILE-1 / 548221168352}  & \textbf{921} & \spspa(1.2)   & 6,118 (1)  \\
     {\scriptsize SHORT-MOBILE-16 / 1566871128} & \textbf{2,085} & \spspa(34.1)   & 3,697 (7)  \\
     {\scriptsize SHORT-SERVER-225 / 5564716}   & \textbf{44,966} & \spspa(28.5)   & 45,794 (1)   \\
     {\scriptsize SHORT-MOBILE-60 / 50044}     & 2,375 & \spspa(27.0)   & 711 (7)   \\
     {\scriptsize LONG-MOBILE-24 / 50044}       &	2,584 & \spspa(35.9)   & 711 (7)  \\
     {\scriptsize SHORT-MOBILE-59 / 50044}      &	3,152 & \spspa(35.83)   & 879 (9)	  \\
\bottomrule
\end{tabular}
\end{center}
  \caption{Number of mispredictions with online and offline sparse linear modeling for branches from Tab.~\ref{tab:motivating_examples}. 
  } \label{tab:online_learning}
\end{table}
%

%
The hardware implementation of SGD-L1 is challenging, since it requires two additional $32$-bit floats per history bit, one for the trainable model weights and one for the cumulative $\ell_1$ penalty. Nonetheless, specific accuracy restrictions, can be used for tracking only a certain subset of static branches with a reasonable entries count and adapt adequately. Similar approaches have already been used during the past in commercial products for learning complicated correlations at runtime, such as Perceptron's training in IBM's z15~\cite{z15_IBM}. 
In \cref{tab:online_learning}, we present the number of mispredictions of SGD-L1 against the offline sparse linear model of \cref{sec:bp_lasso:sparse_modeling}. Within parentheses, we show the average number of non-zero weights that SGD-L1 maintains over its execution. As illustrated, such number never exceeds $36$, and thus, a unit similar to \SLBIU can be efficiently tuned for predicting branches based on the the trained models timely, as we demonstrated in \cref{sec:results:hweval}. Note that online training is not able to learn the sparsity patterns exactly as in the offline setup, with the pleasant exception of the first branch in \cref{tab:online_learning}. However, online sparse modeling does learn an enlarged set of each sparsity pattern. Furthermore, the number of mispredictions are comparable with the ones offline. These early findings suggest that online sparse linear modeling is a promising research direction. 
%

%
\section{Other Related Work}\label{sec:related-work}
%
Evers \etal investigated branch correlations and showed that most of the branches can be predicted efficiently by considering a \emph{selective} history of only a few previous branches. Essentially, our study corroborates the early findings of Evers \etal employing sparse linear modeling to specify the informative branch-history locations. The findings of Evers \etal motivated also the Spotlight branch predictor~\cite{bp:spotlight}. Similarly to our work, Spotlight identifies the important parts of the history offline that are then used by a gshare-like predictor~\cite{mcfarling1993}. During profiling, global-history segments that lead to a branch are analyzed exhaustively to discover the combination that provides higher accuracy. On the contrary, in our work we employ non-exhaustive training methods based on sparse modeling to build a dedicated linear model for predicting each screened branch efficiently by a specialized hardware unit. Fern~\etal~\cite{fern2000dynamic} proposed a decision-tree-based branch predictor with dynamic feature selection to decrease the number of input features, thereby reducing the storage of a tabular branch predictor. Some recent studies~\cite{gupta2021neural,lafiandra2021brat} focused on the implementation of Neural-Network-based predictors that can be trained online, similarly to the online sparse modeling concept we briefly discussed in Section~\ref{sec:online_lasso}. BranchNet~\cite{zangeneh2020branchnet} and the work of Tarsa~\etal~\cite{tarsa2019improving} train offline convolutional neural networks for branches that are hard-to-predict for \tage. Our work differs in the simplicity of the linear models we deploy and in targeting a fundamental control-flow property, the sparsity of branch correlations, that is independent from the predictor's mechanism. Similarly to the above works, however, our model does not target data-dependent branches which have been recently addressed using compiler support by SLB-predictor~\cite{farooq2013slb} or using aggressive run-ahead execution as in Branch Runahead~\cite{pruett2021branch}. 
%
\section{Conclusions \& Future Work}\label{sec:conclusion}
%
This study stimulates the development of sparsity-aware branch prediction for improving accuracy by exploiting a fundamental property of programs control-flow, the \textit{sparsity} of branch correlations. 
We analyzed several traces derived from \specint benchmarks and \cbpfive by capitalizing on sparse modeling methods. Our results demonstrated the existence of numerous sparsely correlated branches. Such branches impede the effectiveness of common branch predictors by putting them under an unnecessary pressure for entries allocation. To eliminate their effects, we propose to employ offline sparse modeling for producing the respective sparse models of branches that will be used for runtime prediction. To that end, we introduce SLBIU, a hardware mechanism specialized to predict branches with offline-prepared sparse models. SLBIU works auxiliary to the primary branch predictor of a CPU design and improves significantly the prediction of sparsely correlated branches. In particular, when combined with \tagesclSmall, SLBIU accounts for up to to 42\% (2.3\% on average) of MPKI improvements with 2\,KB of storage overhead. Furthermore, our circuit-level evaluation with 28nm technology showed that SLBIU is able to deliver predictions in 3 clock cycles at 740\,MHz by requiring no more than 40\,mW of power and as little as 0.34\,mm\textsuperscript{2} of area. Essentially, our results demonstrate the important benefits in branch prediction from identifying and exploiting sparsity effectively. Our study unlocks several other topics for exploration, mainly related to the optimization of offline training of sparse models. In future work, we plan to investigate the effectiveness of other algorithms from the quiver of sparse modeling and also study their runtime adaptability.


\bibliographystyle{IEEEtranS}
\bibliography{references}

\end{document}